\documentclass[10pt,twocolumn]{article}

\usepackage{lmodern}
\usepackage{doi}

\usepackage[margin=0.65in]{geometry}
\usepackage[nocompress]{cite}

\usepackage[T1]{fontenc}
\usepackage[utf8x]{inputenc}
\usepackage[english]{babel}
\usepackage{graphicx}
\usepackage{overpic}
\usepackage{amsmath}
\usepackage{multirow}
\usepackage{rotating}

\renewcommand{\vec}[1]{\mathbf{#1}}

\usepackage{algpseudocode}
\usepackage{algorithm}

\usepackage{hyperref}
\hypersetup{
    colorlinks=true,
    linkcolor=black,
    citecolor=black,
    filecolor=black,
    urlcolor=black,
    linkbordercolor = {1 1 1},
}

\usepackage[square, numbers]{natbib}
\usepackage{pdfpages}

\begin{document}

\title{Towards matching user mobility traces in large-scale datasets}

\author{D\'aniel~Kondor$^{1,2\ast}$, Behrooz~Hashemian$^1$, Yves-Alexandre~de~Montjoye$^3$, Carlo~Ratti$^1$\\
\\
\normalsize{$^1$ Senseable City Laboratory, Massachusetts Institute of Technology, Cambridge, MA, 02139, USA}\\
\normalsize{$^2$ Future Urban Mobility group, SMART, Singapore 138602}\\
\normalsize{$^3$ Dept. of Computing and Data Science Institute, Imperial College London, London SW7 2AZ, UK}\\
\normalsize{$^\ast$To whom correspondence should be addressed; E-mail: dkondor@mit.edu.}
}

\twocolumn[
  \begin{@twocolumnfalse}
    \maketitle
    \begin{abstract}

		The problem of unicity and reidentifiability of records in large-scale databases has been studied in different contexts and approaches,
		with focus on preserving privacy or matching records from different data sources. With an increasing number of service providers nowadays
		routinely collecting location traces of their users on unprecedented scales, there is a pronounced interest in the possibility of matching
		records and datasets based on spatial trajectories. Extending previous work on reidentifiability of spatial data and trajectory matching, we
		present the first large-scale analysis of user matchability in real mobility datasets on realistic scales, i.e.~among two datasets that
		consist of several million people's mobility traces, coming from a mobile network operator and transportation smart card usage. We extract the relevant statistical properties which influence
		the matching process and analyze their impact on the matchability of users. We show that for individuals with 
		typical activity in the transportation system (those making 3-4 trips per day on average), a matching algorithm based on the co-occurrence of
		their activities is expected to achieve a 16.8\% success only after a one-week long observation of their mobility traces, and over 55\%
		after four weeks. We show that the main determinant of matchability is the expected number of co-occurring records in the two datasets.
		Finally, we discuss different scenarios in terms of data collection frequency and give estimates of matchability over time. We show that with higher frequency data collection becoming more common, we can expect much higher success rates in even shorter intervals.	
    \end{abstract}\hspace*{7ex}
  \end{@twocolumnfalse}
]


\section{Introduction}

{N}{owadays} many service providers routinely collect mobility traces of individuals. These constitute various types of data such as call detail records
(CDR) from mobile phone usage~\cite{Blondel2015}, smart cards used in public transportation systems and for identification~%
\cite{Pelletier2011}, financial transactions such as payments made with bank cards or mobile devices~\cite{bbva}, and GPS coordinate updates recorded
by smartphone apps~\cite{Cao2016,crosschecking}. While these provide a valuable data source for researchers~\cite{Lenormand2015,Blondel2015,%
GouletLanglois2016,Toole2015} and also enable various services~\cite{Pelletier2011}, the high amount of
tracking of individual mobility has raised serious concerns about privacy in several different contexts~\cite{Taylor2016,Golder2014}.

This has been emphasized by research that shows that these mobility traces are highly unique, warning that identifying an individual in a mobility
dataset based only on their observed records must be considered as a real possibility~\cite{DeMontjoye2013,DeMontjoye2015}. The basis of this
argument is that since a small number of records uniquely identifies an individual, then reidentification can be achieved based on a relatively small
amount of information, e.g.~by following someone for only a short amount of time, or by merging with an external dataset even with a short timespan.
Furthermore, the possibility of such deanonymization existing \emph{at all} is counter-intuitive to the perception of anonymity achieved in a crowd
of strangers that is typically associated with urban life~\cite{DeMontjoye2013,Milgram1970}.

On the other hand, fusing data at an individual-level from different sources is expected to provide valuable new insights for studies in personal
mobility and urban planning e.g.~by relating mobility and social characteristics~\cite{Lenormand2015}, helping the development of new security and
privacy policies, and benefiting the people involved by offering new services~\cite{Crandall2010,Christen2012}. In accordance with that, previous work has
tried to establish methodology for effectively \emph{matching} mobility datasets based on user traces~\cite{Cecaj2015,Riederer2016,%
Cao2016,Basik2017}.

In this paper, we evaluate \emph{matchability}: the possibility of matching users between large-scale anonymized datasets based on their trajectories. This is in contrast to previous work which has focused on the potential of reidentifiability of anonymized records at the individual level~\cite{Narayanan2008,DeMontjoye2013,DeMontjoye2015}
We utilize two data sources, each of them containing mobility traces of millions of people over the course of one week, a statistically representative
sample of a metropolitan area. While ground truth data about corresponding traces is not available to perform direct evaluations, we address the main
challenges present when dealing with datasets containing fine grain mobility traces of millions of people, as is the case in a major city. Further, we evaluate the expected success rate of a matching procedure
in a realistic scenario, providing first results on such a large scale. Our main contributions in this paper are the following:

	\paragraph*{1} We study the problem of matchability using two datasets which correspond to a significant sample of the population in the area considered.
		To our best knowledge, this is the first attempt to estimate the potential for merging datasets on this scale. This presents a realistic
		scenario in terms of computational complexity and data density, i.e.~the number of false positives is non-negligible.
	\paragraph*{2} We evaluate and develop a matching methodology which can handle data of this size; a main objective is to be able to perform the matching
		without having to evaluate a similarity metric among any pair of users which would present prohibitively high computational complexity. We
		make our implementation available to the research community as open-source software which performs the search  efficiently on datasets consisting of few hundred million records of several million users each.
	\paragraph*{3} We develop an empirical framework for establishing the \emph{matchability} of the datasets and use it to evaluate the expected success rate
		of the matching methodology to estimate the required data collection period for successful matching of users given their activity. This work
		is extensible to more complex search and matching strategies as well.

\section{Related work}
\label{sec_literature}

The problem of \emph{matchability}, along with the related problems of reidentifiability, uniqueness and concerns for privacy has been in the focus of
research for several decades. An early systematic treatment was presented by Fellegi and Sunter in the 1960s~\cite{Fellegi1969}, with the goal of
merging incomplete databases obtained from various sources; their work was inspired by the beginning of the large-scale deployment of computer
systems for storing records in electronic formats and presenting efficient search capabilities for the first time. At that time, the resources
required for acquiring and analyzing large-scale datasets were still limited it to the government, large corporations and a few academic institutions;
privacy concerns were already present, at least with respect to what data could be released to the public e.g.~from census databases~%
\cite{Fellegi1972,Dalenius1977,Dobkin1979,Denning1980}.

Since then, especially the past two
decades brought the proliferation of data which is collected and shared about a significant fraction of the population. A tremendous amount of
information is publicly shared on online social networks (OSNs)~\cite{Mislove2007}, while products and services allowing the tracking of individuals
have gained a high penetration rate. These include credit cards, mobile phones and smartphones, RFID-based payment or identification systems, and
subscription-based online services; all have the possibility of generating a large amount of personal data about their users~\cite{Blondel2015,Pelletier2011,Toole2015,Taylor2016}. While this continues
to raise concerns from several parties, there exist many new opportunities using
insights based on the data, both commercial and academic, giving operators an incentive for sharing the collected data~\cite{Golder2014}.

Given that the amount of data routinely collected about individuals is increasing rapidly, several recent works have tried to evaluate the possibility 
of reidentifying records in a large dataset or matching records from different sources. The work of Sweeney established the notion of $k$-anonymity 
as a guarantee considering the privacy implications of releasing data with records of individuals~\cite{Sweeney2002a,Sweeney2002b}, formalizing 
intuitive requirements that records of individuals should not be unique. Further work showed that $k$-anonymity in many cases can be impractical to 
achieve, especially in the case of high-dimensional but sparse datasets, which typically exhibit a high level of uniqueness. For example Narayanan and Shmatikov, using data released publicly by Netflix, showed that identifying individuals can be possible based on knowledge of only a small number of their records~\cite{Narayanan2008,Narayanan2009}.
Zhang et~al. present a
more general treatment of data disclosures and provide algorithms for obtaining sub-data suitable for releasing under more general constraints than
$k$-anonymity~\cite{Zhang2007}. On the other hand, Dwork argues that for any data release, the possibility of combining it with external information
also needs to be considered and privacy cannot be guaranteed in a general setting; the definition of \emph{differential privacy} is suggested
to quantify the risks of sensitive information that can be gained this way~\cite{Dwork2006}. Their approach and motivation is in many ways similar to
previous work on statistical databases~\cite{Dobkin1979,Denning1980}, with considering more general cases. Further work focused on linking Bitcoin
addresses to IP addresses, showing the limits of privacy~\cite{Biryukov2014,Juhasz2016}, while the problem of record linkage has important applications
in disciplines like biology or astronomy, facilitating probabilistic matching of data obtained from different measurements~\cite{%
Green2006,Budavari2015}.

As location data about individuals is being collected at an increasing pace, several previous studies considered reidentifiability, uniqueness and
matchability among mobility traces of individuals. The work of De~Mulder et~al. was among the first to consider the possibility of identifying anonymized
location traces from mobile networks; looking at a dataset containing data recorded on the phones of 100 volunteers, they achieved an identification
accuracy of over 80\%~\cite{DeMulder2008}. Other works showed that mobility traces of individuals are highly regular and predictable~\cite{Song2010,Scellato2011}, and that this regularity can be exploited for identifying people based on knowledge of their historical trajectories~\cite{Gambs2014}.
In related work, Crandall et~al. look at the problem of inferring
friendships form social media mobility traces based on the assumption that friends will have more shared location updates than strangers~\cite{%
Crandall2010}, while Li et~al. studied the problem of measuring similarity between different people's location histories with the goal of adapting
recommendation systems to include spatial data as well~\cite{Li2008}.

Considering the problem of identifying users based on a sample of their
trajectories in truly large-scale datasets, de Montjoye et~al. introduce the concept of \emph{unicity} and find that mobility traces in mobile phone network usage and credit card transaction data
are highly unique: even in datasets containing the mobility traces of more than a million people, only four randomly picked records from a person's trace uniquely identifies a large
majority of traces~\cite{DeMontjoye2013,DeMontjoye2015}. To mitigate these concerns, He et~al. suggested an advanced anonymization methodology based on
applying the concepts of differential privacy to location data to achieve a form of $k$-anonimity~\cite{He2015}. Nevertheless, these results suggest the possibility of performing a
systematic reidentification among two different large-scale data sources, effectively \emph{matching} traces of users present in both datasets
merely based on their records of movement. In line with this, several studies have performed experiments to match users in distinct
datasets based on trajectories. In an early work, Malin and Sweeney suggest the possibility of deanonymizing genetic data (i.e.~DNA sequences), by
comparing trajectories of patients obtained from medical records with matching DNA sequences they left at several different institutions~\cite{%
Malin2004}. Looking at the problem of matching trajectories in a more typical setting, Cecaj et al.~\cite{Cecaj2015} perform a search for the
trajectories of a sample of about one thousand social media users in a mobile network dataset and present an estimate for the number of users matched
based on a probabilistic model as their data source does not include ground truth information. Riederer et al.~\cite{Riederer2016}
developed a probabilistic matching algorithm based on bipartite graphs on pairs of datasets with readily available ground truth information, while Cao
et al.~\cite{Cao2016} define a signal based similarity measure to integrate data collected via various mobile apps. Recently, Bas\i{}k et al.~\cite{%
Basik2017a,Basik2017} performed a large-scale study testing matching algorithms between CDR, social media and synthetic data.

Most of the above studies considering matching traces in mobility data~\cite{Cecaj2015,Riederer2016,Cao2016} share the limitation that search and
matching is only performed for a limited set of sample users in the range of few tens of thousands; while this provides a test-case for development
of algorithms, the density of the dataset (i.e.~the number of records occurring in a given space and time) will highly affect the methodology applied for matching and the scalablility
and validity of algorithms. Our datasets allow us to study the problem of matchability on the scale of millions of users. 
As this presents new challenges in terms of computational performance, we only consider a matching strategy which performs an
efficient search instead of evaluating any possible pair of users as a match candidate. However, given the scale of the dataset, we will have a realistic estimate on the
probabilities of finding false positive matches, highly unlikely when using a dataset containing only a few thousand trajectories.
In this respect, our work is most similar to Bas\i{}k et al.~\cite{Basik2017,Basik2017a}, who utilize two large datasets as well (containing of several hundred
thousand and a few million users respectively), but do not present relevant statistics about the matches between these two datasets due to the lack
of ground truth data. They do present an extensive analysis on the quality of matching between a synthetic dataset generated from CDR
data and the original CDR dataset. Using a different approach, we now present a probabilistic framework to assess the possibility of matching
real-world mobility traces between two large-scale datasets.

\section{Data description}
\label{sec_data}

\begin{figure*}
\centering
		\includegraphics{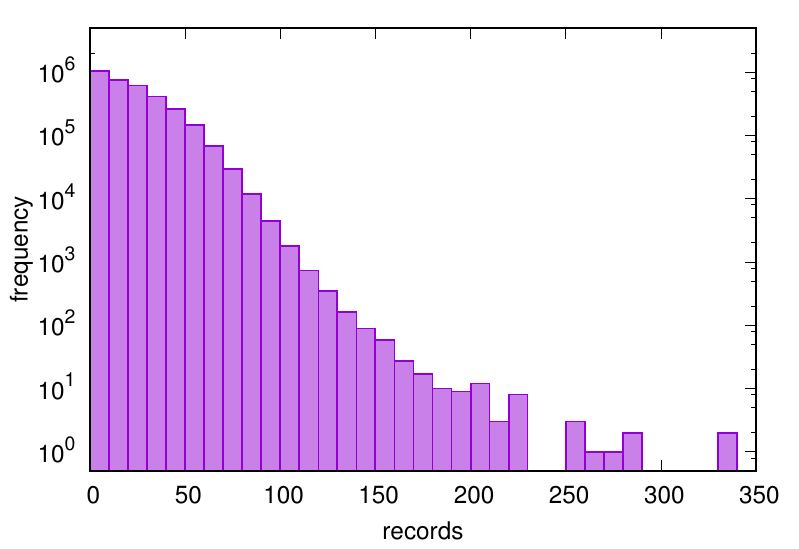}
		\begin{overpic}{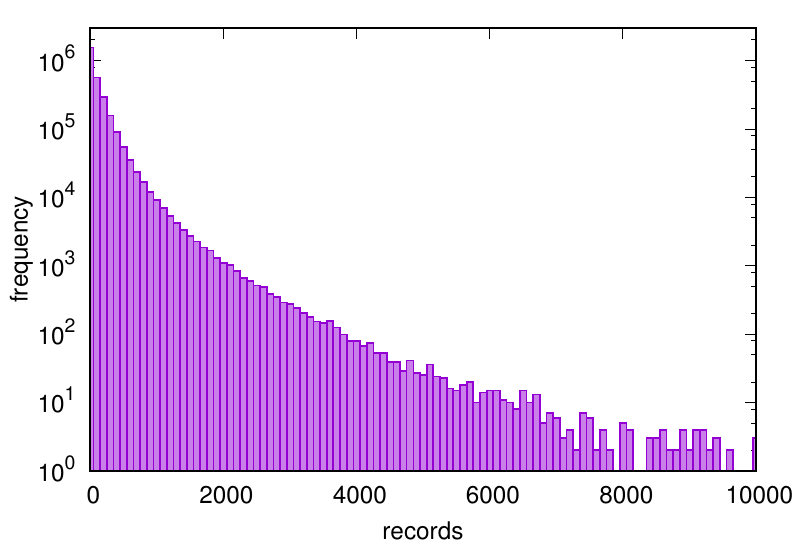}
			\put(39,23){\includegraphics{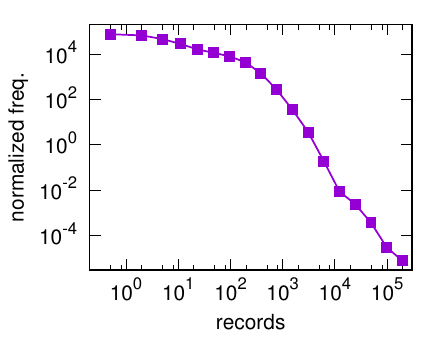}}
		\end{overpic}
		\caption{{\bf Distribution of number records per user.} Left: number of taps per user in the transportation dataset (the histogram was
			calculated with a bin size of $10$). Right: number of records per user in the mobile communication dataset (the histogram was calculated
			with a bin size of $100$); the inset shows the same distribution with logarithmic sized bins and a logarithmic $x$-axis to better show
			the tail of the distribution. We expect that ``users'' with extremely high number of records actually correspond to automated services.}
		\label{checkins}
\end{figure*}

In this work, we utilize one week of mobile communication and transportation data from the city state of Singapore recorded during the spring of 2011.
The mobile communication dataset was provided by Singtel, the largest mobile network operator (with a market share of over 45~\%) and contains
485,237,708 individual records of 2,844,721 users, where one record represents the start or the end of a call (either placed or answered) or sending
or receiving a text message and includes the timestamp and the geographic coordinates of the antenna the user was connected at the time. The
transportation data comes from the Singapore Land Transportation Authority (LTA) and is based on the smart cards used by the electronic fare system
on buses and trains. This dataset contains 71,319,524 individual records produced by 3,348,628 unique smart cards where one record corresponds to
either boarding or exiting a bus or train and includes the timestamp and the coordinates of the corresponding stop. Train rides always include both
the start and the end of the journey with the possibility of transfers in between not necessarily recorded, i.e.~the start and end stations can be
on different lines. Bus rides include the end of the journey only when the passenger performs an additional \emph{tap-out} while exiting
the vehicle. Doing so is optional and incentivized by providing fare discounts (i.e.~the fare is billed based on the actual travel distance instead of a flat fee);
this is highly effective as evident by the fact that about 94 \% of bus rides in our dataset record the end of the trip as well.

\begin{figure*}
	\includegraphics{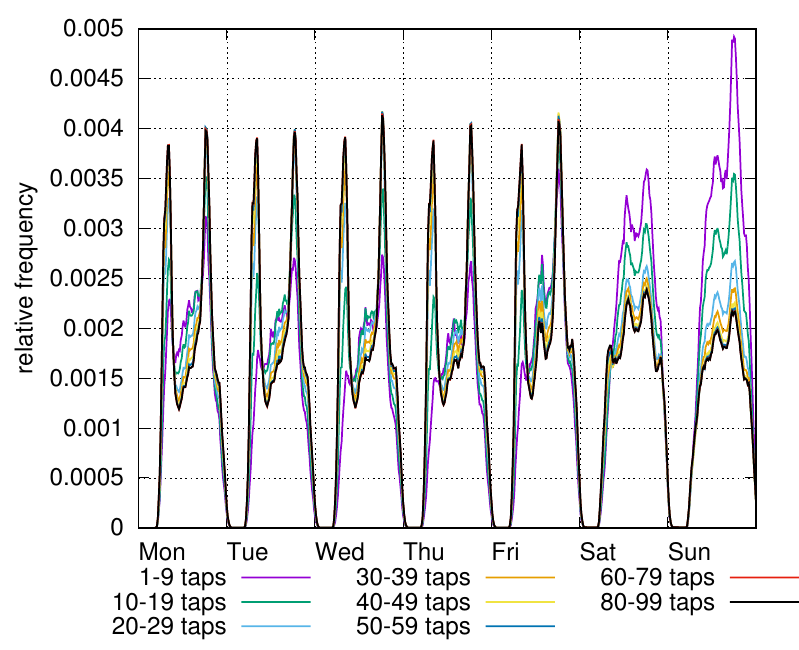} \quad
	\includegraphics{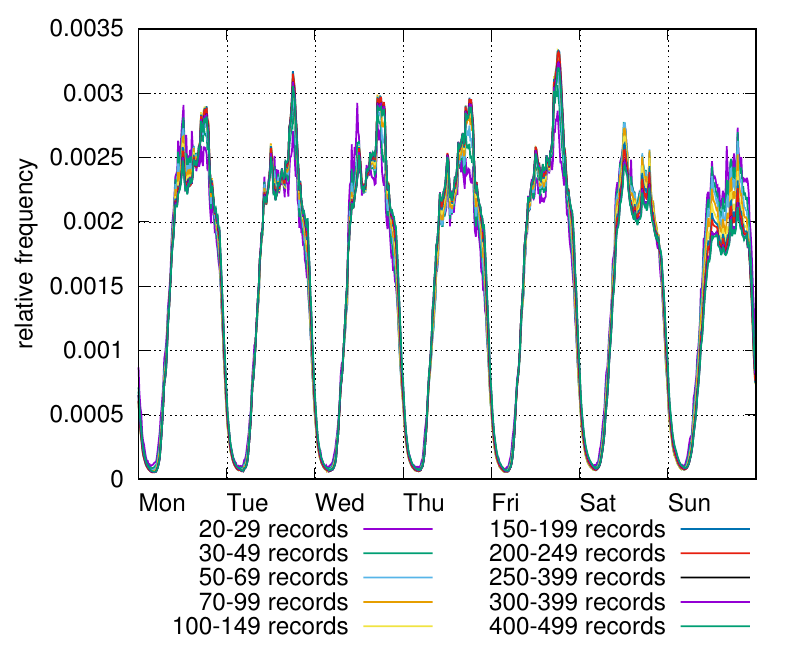}
	\caption{{\bf Temporal distribution of activities}. Left: users in the public transportation dataset; right: users in the mobile communication
		dataset. Activities were aggregated to 15 minute intervals for the purpose of generating this figure. We see that temporal distribution of
		activities is quite similar for all groups in the mobile network data, while there are significant differences in the case of the
		transportation dataset. We expect that the probability of finding matching records from the same user in the two datasets to depend whether the user is active at the same time in both of them; thus difference of the distribution of activities will be relevant when comparing the distribution of records of temporal matches between the
		different groups of users.}
	\label{tdist_lta}
	\label{tdist_singtel}
\end{figure*}

\begin{figure}
	\begin{overpic}{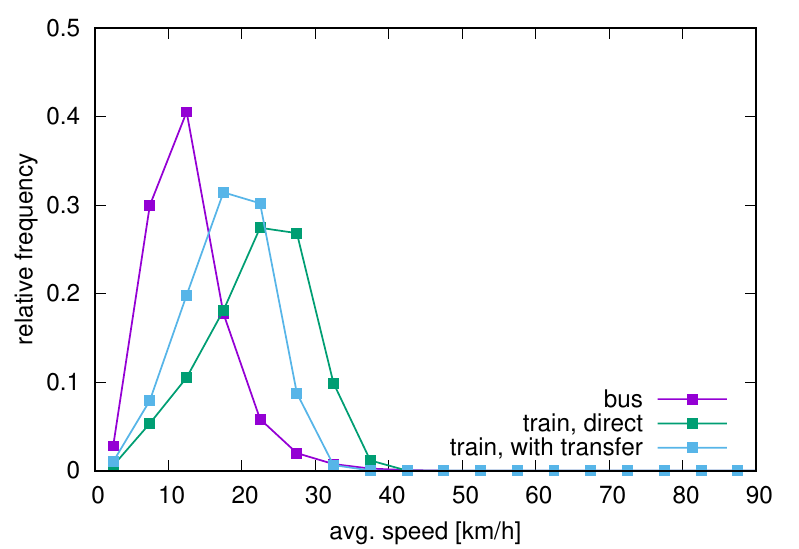}
		\put(38,23){\includegraphics{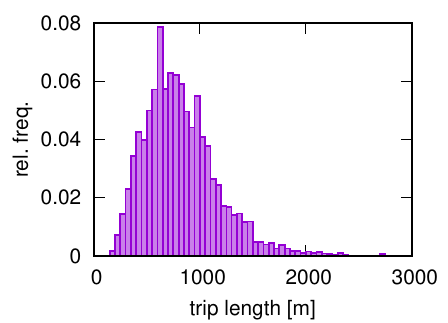}}
	\end{overpic}
	\caption{{\bf Distribution of average trip speeds} in the transportation dataset. Distances were calculated as shortest geographic distance
		between the start and end stops. This implies that the speed of the vehicles will be probably higher, but since we are interested in the
		typical radius people travel during a certain time period, this estimation aligns with our notion of spatial neighborhoods in
		Section~\ref{sec_spmatching}. For all modes of transportation, typical speeds are below $40\,\mathrm{km}/\mathrm{h}$. Larger average speeds
		constitute only $0.13\%$ of bus rides and $0.08\%$ of train rides which are likely to be measurement errors in the dataset. The inset shows the
		distance spanned by trips shorter than five minutes and supports our notion of using a $2\,\mathrm{km}$ spatial search radius for people
		taking transit (see Section~\ref{sec_spmatching}).}
	\label{lta_speeds}
\end{figure}

As a first step to characterize the data, we look at the temporal distribution of records as well as basic statistics of the number of records per user (summarized in Fig.~\ref{checkins}). It is
clear that both data sources are relatively low-density as especially in the case of the transportation dataset, most people are expected to travel
by public transportation only a few times per day and we only have records at the start and end of their trips.

To better characterize temporal distributions of activity during the week, in Fig.~\ref{tdist_lta} we display activity distributions for transportation and mobile network users. We do this by grouping users based on their level of activity. The distributions are remarkably similar among the groups of users with different level of activity for the
mobile network dataset, but they are significantly different in the transportation dataset. This can be explained by noting
that the less active user groups are more likely to be casual users, i.e.~people who do not use public transportation as their primary means for
commuting to work; this seems to be the largest difference between the groups. Also, in Fig.~\ref{lta_speeds}
we display the distribution of average travel speeds in the transportation dataset, used later as the basis of defining spatiotemporal neighborhoods
of transportation events.

\subsection{Evaluating unicity}

\begin{figure}
	\includegraphics{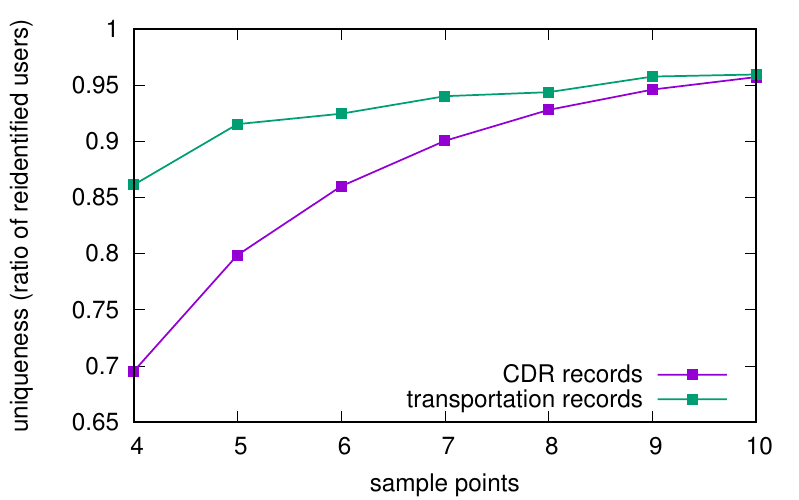}
	\caption{Evaluating unicity with $d = 500\,\mathrm{m}$ and $\tau = 5\,\mathrm{min}$.}
	\label{matchself1}
\end{figure}

Before proceeding to estimate the matchability of the two datasets, we first evaluate the uniqueness of mobility traces using unicity as the measure~\cite{DeMontjoye2013}. We select random samples from users' records and test whether these uniquely identify them in 
the dataset. We show results for \emph{unicity} (i.e.~the ratio of users uniquely identified) in Fig.~\ref{matchself1}. These are similar to 
that of previous work~\cite{DeMontjoye2013, Cecaj2015}, although slightly lower. This is possibly the result of our 
datasets being denser (both in space and in time and also in the case of public transportation use, we expect the train stations to be more crowded). Furthermore, in the case of the communication dataset, instead of grouping records by antennas, we evaluate spatial proximity based on the Voronoi-polygons centered on antenna locations. This is a difference from previous studies~\cite{DeMontjoye2013, Cecaj2015} which we expect to result in lower values of unicity, but model the process of matching records in two different datasets better as well. We point out a fundamental 
difference from matching two datasets: when evaluating unicity, we know that there is a match for each record (i.e.~the record itself), and 
then increasing the search radius (either in space or time) adds potential false positive matches, giving rise to decrease in unicity~%
\cite{DeMontjoye2013,DeMontjoye2015}. On the other hand, when matching records from different datasets, most of the records are likely to not have a
match in the other dataset from the same user (see also Fig.~\ref{tmatchlta4049} and the related discussion in Section~\ref{sec_validating});
increasing the search radius will increase both the chance of a record of the same user to be matched and the number of false positives.

\section{Spatiotemporal matching}
\label{sec_matchingmethod}

\begin{figure}[t]
	\includegraphics[width=0.5\textwidth]{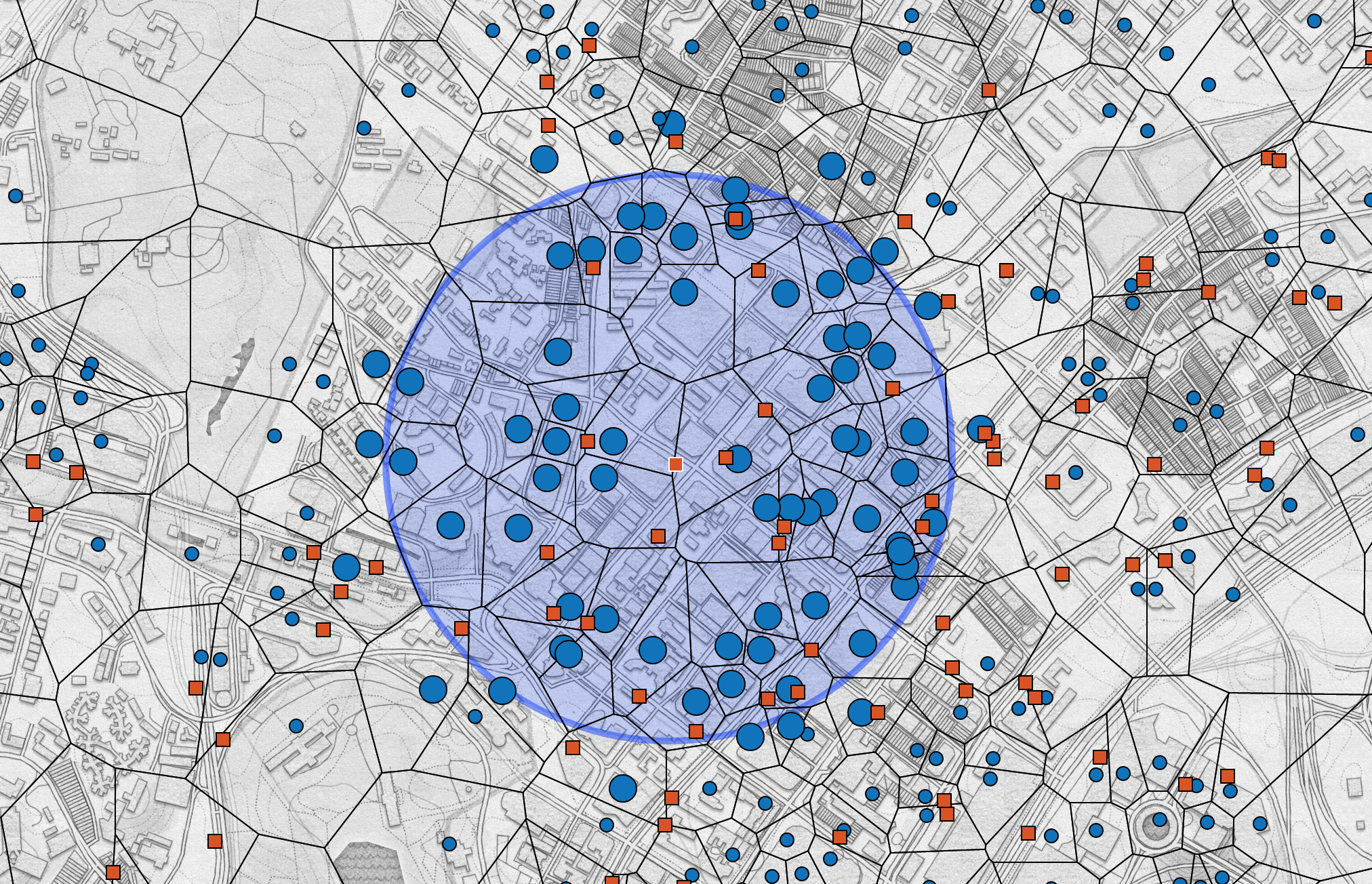}
	\caption{{\bf Voronoi tessellation of cellphone antennas.} Red rectangles show public transportation stations and blue circles show antenna
		locations. The transparent circle is centered at one public transportation station in the middle and antennas that can be considered as
		possible matches are highlighted as larger blue circles. Notice that this includes antennas whose locations are outside the blue circle as
		their corresponding Voronoi-polygons still intersect with it. Note that this figure is only for illustrative purposes and thus does not
		display the real antenna locations used in our study, which are considered confidential information by the data provider. The antenna
		locations displayed here were obtained from \texttt{https://opencellid.org/} for the sake of this illustration.}
	\label{voronoi_match}
\end{figure}

The main goal of this work is thus to study
matchability as a function of the length of the data collection period and the amount of data available per individual.
We emphasize that to do this reliably, it is necessary to have a
realistic estimate of the levels of activity both in terms of spatiotemporal distribution and density. While our datasets do not include ground truth information on matching trajectories, they provide a significant sample of all activities in a compact metropolitan area, allowing to do such estimation more reliably. However, having such datasets presents obvious challenges in
terms of data handling and computational complexity, limiting methodology to those that scale up efficiently to these large scales.

To be able to evaluate matchability, we first need to define the matching process whose expected success rate we will estimate. In this section, we present a
simple choice for matching users where expected matching performance can be evaluated probabilistically, without the need for ground truth data.

Our methodology is a spatiotemporal search which yields candidate user pairs and can be carried out efficiently using standard indexing
techniques without having to perform a comparison of all possible pairs of trajectories. We search for matching points in the spatiotemporal
neighborhood of each record of every user, excluding candidates who have records temporally close, but spatially distant. For each user we select the one in the other dataset with the highest number of matching
records as a candidate.

While our matching procedure is relatively simple, it scales well to datasets of several hundred million points each
and can be easily adapted to estimate matchability based on the probability of finding false positive matches.
It would also allow further extensions (e.g.~with weighting matches based on local density) if ground truth data were available to perform training on. In this sense, we expect our results on matchability
to be lower bounds on the success rate of any matching algorithm developed with real-world data.

\subsection{Notations and preliminaries for spatiotemporal matching}
\label{sec_spmatching}

We first need to define when we consider two records to be matching. Let $C$ and $T$ denote two mobility datasets (i.e.~communication and
transportation), with $n_C$ and $n_T$ individual users respectively. We denote the records of user $i$ in either dataset as $x_{ik}^{\alpha} =
( \vec{x}_{ik}^\alpha, t_{ik}^\alpha )$, where $\alpha = C,T$, $i = 1,2,\ldots n_\alpha$ and $k = 1,2,\ldots r_i^\alpha$ where $r_i^\alpha$ is the number
of records of user $i$ in dataset $\alpha$
In the case of the transportation dataset ($\alpha = T$), for each record, we will further use a flag
$S_{ik}^T = 0,1$ which indicates whether that the record corresponds to the start ($S_{ik}^T = 1$) or the end ($S_{ik}^T = 0$) of a journey. Using
these notations, we define two points to be a \emph{spatial match} if they are close in space and time:
\begin{equation}
	| \vec{x}_{ik}^C - \vec{x}_{jl}^T | \leq d \quad \textrm{and} \quad | t_{ik}^C - t_{jl}^T | < \tau
\end{equation}
for some parameters $d$ and $\tau$ which define spatiotemporal neighborhoods. Further, we define two points to be an \emph{impossible match}, if they
are close in time, but separated in space:
\begin{equation}
	| \vec{x}_{ik}^C - \vec{x}_{jl}^T | > d \quad \textrm{and} \quad | t_{ik}^C - t_{jl}^T | < \tau \, \textrm{.}
\end{equation}
Following Bas\i{}k et~al., we also use the term \emph{alibi} to refer to impossible matches~\cite{Basik2017}. We use the term \emph{temporal match} for pairs of points which are either a spatial or impossible match (i.e.~any pair of points separated by less than $\tau$ time irregardless of their spatial distance). Essentially these
definitions mean that we consider two points to possibly belong to the same user if they are separated by maximum $\tau$ in time and $d$ in space,
while we consider them to certainly belong to distinct users if the temporal separation is less than $\tau$ and the spatial separation is more than
$d$. Note that pairs of points with temporal separation larger than $\tau$ are not considered in any way.

To perform dataset matching, we then need to choose the $d$ and $\tau$ parameters such that they are consistent with typical mobility patterns of
individuals. To better accommodate for the characteristics of urban movements, we further refine these thresholds by differentiating between walking
and traveling with transit, based on the transportation records. We thus use separate parameters $d_w$, $\tau_w$ for walking and $d_t$, $\tau_t$ for
transit. We refine the definition of the parameters used for establishing temporal, possible and impossible matches as following:
\begin{equation}
	\begin{array}{l|c||l|l}
		S_{ik}^T & t \, \textrm{order} \, & \, \, \tau & d \\ \hline \hline
		1 \, \textrm{(start)} & t_{ik}^C < t_{jl}^T \,\, & \,\, \tau_w & d_w \\ \hline
		0 \, \textrm{(end)}   & t_{ik}^C > t_{jl}^T \,\, & \,\, \tau_w & d_w \\ \hline
		1 \, \textrm{(start)} & t_{ik}^C > t_{jl}^T \,\, & \,\, \tau_t & d_t \\ \hline
		0 \, \textrm{(end)}   & t_{ik}^C < t_{jl}^T \,\, & \,\, \tau_t & d_t
	\end{array}
\end{equation}
In practice, we chose the parameters as $d_w = 500\,\mathrm{m}$, $\tau_w = 10\,\mathrm{min}$, $d_t = 2\,\mathrm{km}$ and $\tau_t = 5\,\mathrm{min}$
according to the typical travel speeds we found in the data. Note that this implies a typical average transit velocity of $24\,\mathrm{km}/\mathrm{h}$
during this $5\,\mathrm{min}$ period; for bus rides, typical average travel speeds are below this (see Fig.~S4 in the Supplementary Material),
while for train rides, one has to consider that this time interval includes the time needed to reach and enter or exit the train from the station
entrance where the smart card is validated. Looking at the distribution of distances spanned by trips shorter than $5$ minutes (as shown in the inset
of Fig.~S4 in the Supplementary Material)
and between $5$ and $10$ minutes, we estimate that only $1\%$ of trips shorter than 5 minutes has a distance larger than $2\,\mathrm{km}$, while only
$2\%$ of trips between 5 and 10 minutes spans a distance larger than $4\,\mathrm{km}$, meaning that our choice
of spatiotemporal neighborhoods is realistic for a large majority of trips.

We finally need to take into account the spatial uncertainties of the data. While in the case of transportation records, the location of stops
is exact (i.e.~a record implies that the corresponding user was present at the exact location at that exact
time), for the cell phone data, antenna locations are only an approximation of the users' location as they could be anywhere in the antenna's corresponding reception
area or even in a neighboring region if the antenna closest to them is experiencing heavy traffic. To take this into account, we calculate the Voronoi tessellation of unique antenna locations, and consider the user to be possibly present
anywhere within the Voronoi cell which corresponds to a particular antenna. For a certain public transportation stop and cell
phone antenna, records at these two which match temporally will be considered as \emph{spatial matches} if a circle of $d$ radius around the
transportation stop and the Voronoi-polygon associated to the antenna have any overlap, while these records will be considered \emph{impossible
matches} if there is no overlap. We display an example of evaluating such overlaps in Fig.~\ref{voronoi_match}.

\subsection{Matching users}

\begin{figure*}
\centering
	\includegraphics{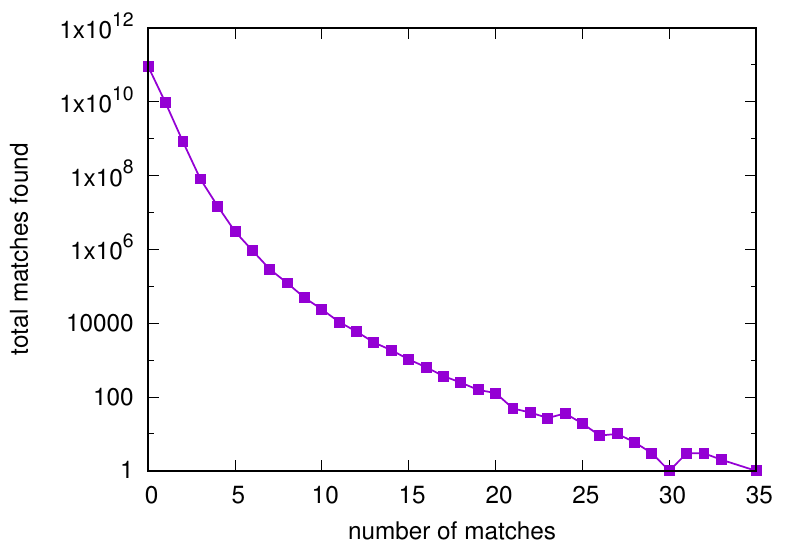}
	\begin{overpic}{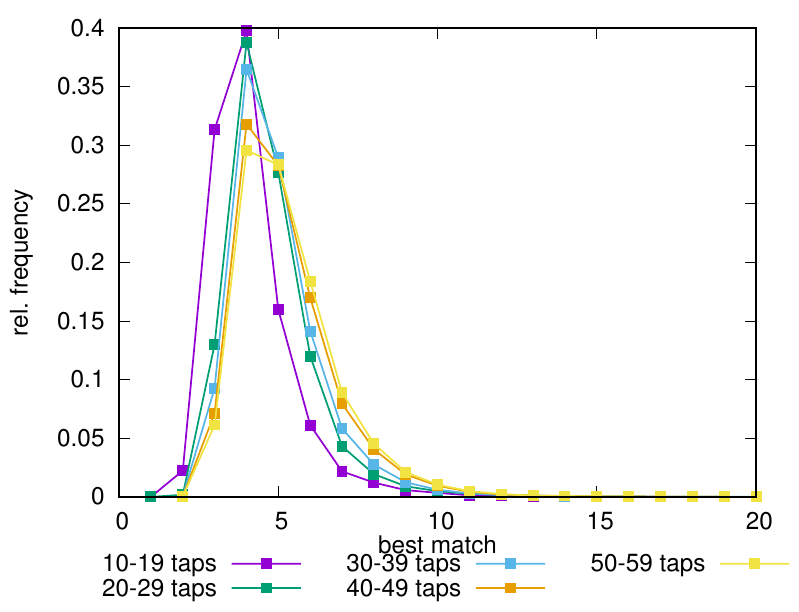}
		\put(45,23){\includegraphics{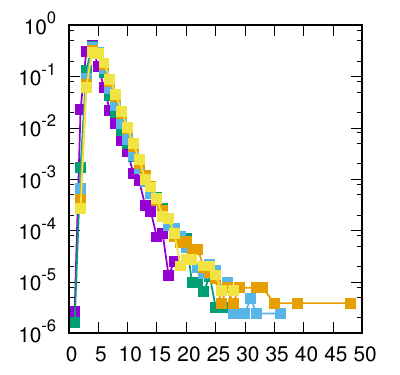}}
	\end{overpic}
	\caption{{\bf Distribution of spatial matches.} Left: distribution of the number of spatial matches found between any two users in the dataset.
		Alibis are included as zero matches
		in accordance with the definition of $P_s$ in Section~\ref{sec_validating}). Right: Distribution of the best spatial match (i.e.~user pairings with the maximum number of matches for each user) when
		selecting transportation users and searching for matching users in the CDR data, grouped by activity. Each distribution is normalized among the activity group. The inset shows the same distribution on a logarithmic scale.}
	\label{matchhistall}
\end{figure*}

Based on the previous considerations, we perform a search procedure among the two datasets which results in a list of candidate matching pairs. While
the size of the data is fairly large, we exploit the fact that there is a limited number of possible matching mobile network antenna -- transportation stop
pairs, pre-compute the list of these based on the Voronoi-polygons (see the previous sections) and use an indexing in time and by the antenna or
stop IDs. This allows us to avoid performing a spatial search and instead use a range search (in time) along with a dictionary search (among the possible
antenna -- stop pairings). Using this strategy, the search for possible match candidates can be performed in the matter of 40 hours using a mid-grade
server with 18 virtual cores and 96~GB available memory. We estimate that evaluating all temporal matches for each user pairs would take approximately 12 days.
Note that this latter computation requires considering all $n_{C} \times n_{T}$ user pairs, while during the matching procedure, this complexity is
significantly reduced by only considering candidate pairs that have co-occurring data points, resulting in the shorter computational
time. We make the source code of all programs utilized in these calculations available at \texttt{https://github.com/dkondor/matching}. A formal overview of the matching procedure is presented in Algorithms~\ref{alg1} and~\ref{alg2}.

\begin{algorithm}[t!]
	\begin{algorithmic}
		\State $T = \{i,\{x_{ik}^{T}\}_{k=1}^{r_{i}^{T}}\}_{i=1}^{n_{T}}$ transportation dataset
		\State $C = \{i,\{x_{ik}^{C}\}_{k=1}^{r_{i}^{C}}\}_{i=1}^{n_{C}}$ communication dataset
		\State $M = $ \{empty set of match candidates\}
		\ForAll{$i \in \{1,2,\ldots n_{T}\}$} 
			\State $A = $ \{empty set of users from $C$ with alibis\}
			\ForAll{$k \in \{1,2,\ldots r_{i}^{T}\}$}
				\State $S = $ search for spatial matches around $x_{ik}$
				\State $U_{C} = $ distinct users from $S$
				\ForAll{$j \in U_{C}$}
					\If{$j \not\in A$ and $(i,j) \not\in M$}
						\State $m = \texttt{CompareUsers}(i,j)$
						\If{$m$ is alibi}
							\State add $j$ to $A$
						\Else
							\State add $(i,j,m)$ to $M$
						\EndIf
					\EndIf
				\EndFor
			\EndFor
			\State optionally: only keep top few matches of $i$ in $M$
		\EndFor
	\end{algorithmic}
	\caption{\small Basic algorithm to calculate matching user pairs. The input is the two datasets, sorted by user ID and time. The second dataset
		(the communication dataset $C$ in this case) is also stored in an index which allows quickly finding points spatiotemporally close to
		a given query, described in the main text in more detail. The main loop selects candidate matching pairs based on spatiotemporal proximity
		which are then evaluated using the more detailed \texttt{CompareUsers} function, listed separately as Algorithm~\ref{alg2}.
		Performing this comparison in this
		separate function is necessary so that we can ensure that each point from either trajectory is considered at most once as a match. Impossible
		matches for each user are kept track in the $A$ set, while potential matches are kept track in the $M$ result set. The $M$ set can be pruned after
		processing each user to only retain the top match (or top few matches), limiting the size of the output. Computational complexity
		scales with the number of candidate point pairs found, but the $M$ and $A$ sets ensure that the \texttt{CompareUsers} function is called at most once for each candidate user pair. As the spatial search is implemented by a simple index lookup, its complexity scales with the total number of points found.}
	\label{alg1}
\end{algorithm}

\begin{algorithm}
    \begin{algorithmic}
        \Function{CompareUsers}{$i$,$j$}
            \State $x_{ik}^T$ records of user $i$ in $T$
            \State $x_{jk}^C$ records of user $j$ in $C$
            \State ensure that $x_{ik}^T$ and $x_{jk}^C$ are sorted by time
            \State $M = \{$ empty set for matched points from $C$ $\}$
            \State $m = 0$ result: number of matches
            \ForAll{$k \in \{1,2,\ldots r_i^T\}$}
                \State $m_k = \texttt{False}$
                \ForAll{$l$ s.t. $x_{ik}^T$ and $x_{jl}^C$ are a temporal match}
                    \If{$x_{ik}^T$ and $x_{jl}^C$ are spatially inconsistent}
                        \State \Return alibi
                    \ElsIf{$m_k = \texttt{False}$ and $x_{jl}^C \not\in M$}
                        \State $m = m + 1$
                        \State $m_k = \texttt{True}$
                        \State add $x_{jl}^C$ to $M$
                    \EndIf
                \EndFor
            \EndFor
            \State \Return m
        \EndFunction
    \end{algorithmic}
    \caption{The \texttt{CompareUsers} function from Algorithm~\ref{alg1}. This function checks that the trajectories of users $i \in T$ and $j \in C$ are compatible (contain no alibis) and counts the number of matching points. It employs the constraint that each point is counted at most once as a match. This is achieved by iterating over the points in time order in both trajectories and keeping track of which points have been already recorded as part of a match. The computational complexity scales with the number of temporal matches between the two trajectories, as all of these need to be checked for spatial consistency. Note that in practice, this function will receive records that are already sorted by time, so a sort step is not necessary and is included only for the clarity of presentation.}
    \label{alg2}
\end{algorithm}

We consider a pair of users $i \in T$ and $j \in C$ as a candidate if any of their record pairs $x_{ik}^T$ and $x_{jl}^C$ are spatial matches
and none of their possible record pairs are alibis. In this case, we define the number of matches $m_{ij}$ between the two users as the maximum
number of possible match pairs such that each record is used once at maximum (i.e.~we exclude multiple matches for a single record). In the case of
any alibis, we define $m_{ij} \equiv 0$. For each transportation user, we select the CDR user
with the highest number of matches as a candidate to be its counterpart. We then refer to such user pairs as a \emph{pairing} to distinguish in language between the case of matching points and matching trajectories. We display the distribution of the number of matches found in Fig.~%
\ref{matchhistall}.
We note that while the importance of using alibis can be easily understood, not all previous work have utilized it.
The probabilistic models in studies~\cite{Riederer2016,Cao2016} only deal with co-occurring events, disregarding the possibility of using such
negative evidence to prune candidate matches; on the other hand the work of Cecaj et al.~\cite{Cecaj2015} perform a similar filtering (using the term
``exclusion condition'' for it), while Bas\i{}k et al.~\cite{Basik2017} define the term alibi, explicitly test for the importance of such filtering and find that doing so significantly increases the quality of matching.

\section{Estimating matchability}
\label{sec_validating}

Having presented a matching methodology, we proceed with estimating its expected rate of success based on considerations of the statistical
properties of matches.
The main question we evaluate in the rest of the paper is whether a user pairing produced by the matching algorithm really corresponds to trajectories of the same person or if it
is a false positive, i.e.~two people who happened to appear in the dataset together $m_{ij}$ times at random.

We proceed in three steps: first, we estimate the probability distribution of having a certain number of temporal matches between the records of any
two users in the two datasets. Second, we use this as a basis for estimating the probability of an individual having 
a certain number of true positive matches among their two trajectories in the two datasets. Finally, we use the observed distribution of spatial matches to estimate the
probability of obtaining more false positive matches randomly than the two number of positive matches estimated from the temporal match distribution.

\subsection{Preliminary assumptions}

We begin by defining probability distributions for obtaining a specific number of matches among user pairs from the two datasets and then estimating
these from our data. We use the following notations:

\begin{itemize}
	\item $P_t (m | i,j)$ is the probability distribution of users $i$ and $j$ having exactly $m$ temporal matches in the data.
	\item $P_s (m | i,j)$ is the probability distribution of users $i$ and $j$ having exactly $m$ spatial matches in the data
		and no impossible matches. We define $P_s (0 | i,j)$ to include both the case when the two users have zero temporal matches and the case
		when they have $>0$ temporal matches of which at least one is an alibi, i.e.~spatially inconsistent.
	\item $P_s (m | i)$ is the probability distribution of user $i$ having exactly $m$ possible matches with \emph{any} user in the other dataset. 
\end{itemize}

\begin{figure*}
	\centering
	\begin{overpic}{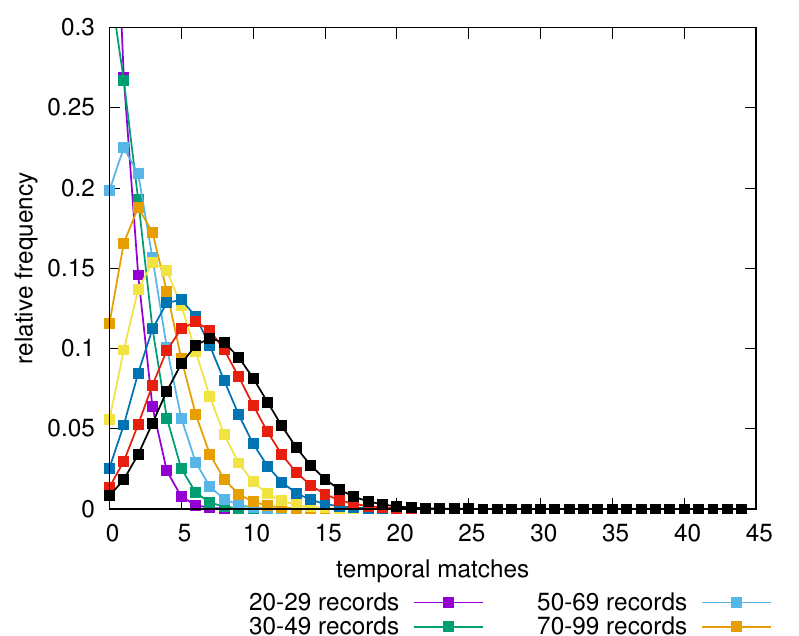}
		\put(35,32){\includegraphics{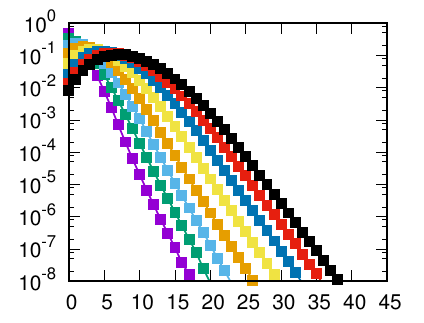}}
	\end{overpic}  
	\begin{overpic}{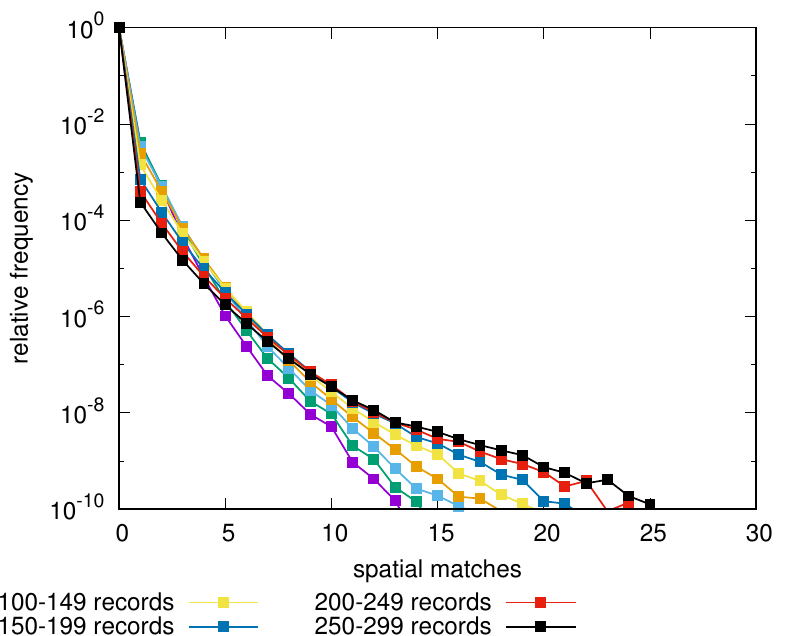}
	\end{overpic}
	\caption{{\bf Probability distributions for matches} compiled for LTA users with activity between 40 and 49 taps and Singtel users grouped by
		the number of calls. Left: distributions of temporal matches, i.e.~$P_t (m | r_i, r_j)$, which we use to estimate the probability of a
		person having a certain number of real matches among their records in the two datasets; the inset shows the same distributions with a
		logarithmic scale. Right: distributions of possible matches, i.e.~$P_s (m | r_i, r_j)$, which we use to estimate the probability of getting
		false positive matches.
		}
	\label{spratios}
	\label{tmatchlta4049}
\end{figure*}

We note that for known trajectories, the exact number of matches can be calculated. In our case however, we do not know which trajectory belongs to which person, thus we will use empirical estimates of these distributions based on the counts of matches among groups of trajectories. To be able to do so and use these distributions to calculate reidentification probabilities we employ some assumptions. First, we
assume that $P_t (m | i,j)$ depends neither on whether users $i \in T $ and $j \in C$ represent the same person, nor on
whether the two trajectories in the data are spatially consistent (i.e.~they could be alibis); thus we can use an empirical estimation of it based on the data without the need
for ground truth on user pairings. We note that this assumption means that when considering the trajectories of the same person in the two datasets, $P_s (m | i,i) = P_t (m | i,i)$, as in this case,
all matches are spatially consistent. Second, we group users together by activity and estimate $P_s$ and $P_t$ for each groups empirically, i.e.~$P_t (m | i,j) = P_t (m | r_i,r_j)$, where $r_i$ and $r_j$ are the number of records we have about them in the dataset. This includes the assumption that matches are independently and identically distributed among any pair of users in these groups.
To improve the statistics and limit the complexity, we use moderate-sized subgroups of user activities instead of
estimating the distributions for every possible $(r_i, r_j)$ pair. As an example, we display the obtained $P_t$ and $P_s$ distributions for
transportation users between 40 and 49 records (between 20 and 25 trips) and several groups of mobile network users in Fig.~\ref{tmatchlta4049}.
The $P_t$ distribution of temporal matches shows that our dataset has the limitation that for the most typical combinations (3
trips per day and 5-10 calls or texts per day, resulting in about 42 taps and 35-70 CDR records in our data), the expected number of matches is still relatively small (1-5). As expected, the $P_s$ distribution decreases rapidly as well, based on the high unicity in the
data. We display average distributions (i.e.~for transportation users between 40 and 49 records and any mobile network user) in Fig.~\ref{matchescmp}. We further display the observed ratio of $P_s / P_t$, which we can interpret as the ratio of probabilities having a given number of spatially consistent matches when considering the trajectories of different people vs considering the trajectories of the same person.

\subsection{Matchability estimate}

\begin{figure}
\centering
	\includegraphics{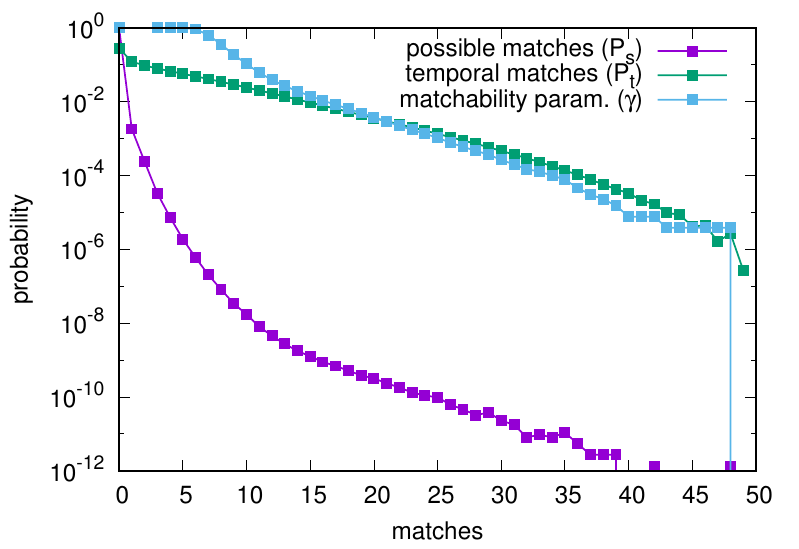}
	\caption{{\bf Estimating expected success ratios.} Aggregated probability distributions of matches for transportation users with activity between
		40 and 49 taps and any mobile network user, displayed along with the calculated matchability parameters calculated from the possible match
		distributions.}
	\label{matchescmp}
\end{figure}

\begin{figure}
	\includegraphics{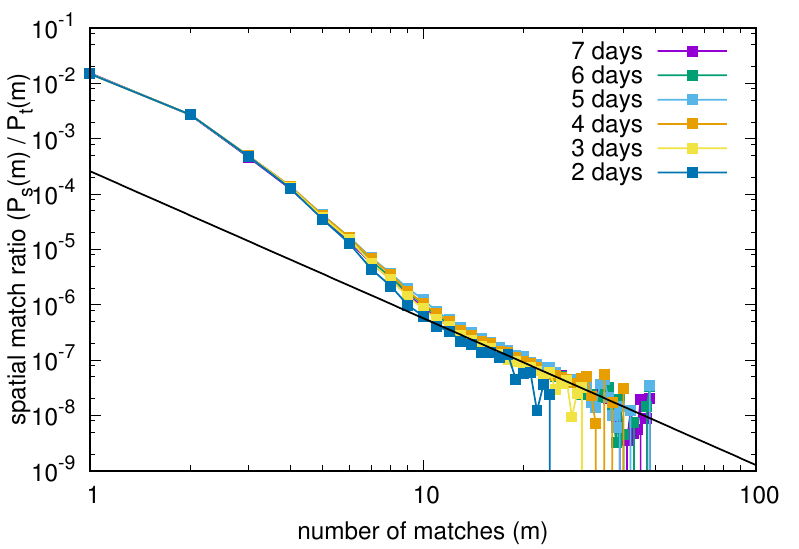}
	\caption{{\bf Ratio of spatially consistent matches to temporal matches}, i.e.~$P_s(m) / P_t(m)$. The data in this figure was created after
		including all user activities in our dataset, representing the distributions to be found among any pair of records of user trajectories. The
		different lines correspond to different subsets of the data; we can see that all of these behave quite similarly, supporting the assumption
		that the success ratios will also behave similarly regardless of the time period considered.  The black
		line shows a power-law fit to the last section. The fitted function is given as $P_s (m) / P_t (m) = 8 \cdot 10^{-6} m^{-1.52}$.}
	\label{spratios2}
\end{figure}

Using our previous assumptions, we estimate the probability of successfully reidentifying a user with given activities $(r_1, r_2)$ in the two datasets. In
accordance with the previous assumptions, we use $P_t (m | r_1,r_2)$ for estimating the probabilities of getting a certain number of real matches
among the two traces: we thus assume that the real number of matches is $m^*$ which is drawn from the probability distribution $P_t$.  We then assume
that the reidentification is successful if there is no other user with possible matches of $m^*$ or more occurring randomly.
Using our previous definitions, the possibility of this for one user is given by $\sum_{m' \geq m^*} P_s (m' | r_1)$. As this can occur for each user (a total of
$n_C \approx 2.84M$ in our communication dataset), we need to calculate the probability that none of them has such a match, given by
$\gamma (m^* | r_1) \equiv \left ( 1 - \sum_{m' \geq m^*} P_s (m' | r_1) \right )^{n_C}$ which we denote as the \emph{matchability parameter} for a
given user activity $r_i$ and number of matches $m^*$. Since we consider the number of real matches ($m^*$) as a random variable as well, we then
proceed by calculating the probability of successful reidentification, denoted by $p_x (r_1, r_2)$ as the expected value of $\gamma$:

\begin{multline}
	p_x (r_1, r_2) = \langle \gamma (m | r_1) \rangle = \sum_m P_t (m | r_1, r_2) \gamma (m | r_1) = \\
		= \sum_m P_t (m | r_1, r_2) \left ( 1 - \sum_{m' \geq m} P_s (m' | r_1) \right )^{n_C}
	\label{eq_px}
\end{multline}

We note that in practice, we do not know the real values of the $(r_1, r_2)$ pair of activities. Instead, we calculate $p_x (r_1, r_2)$ for 
different groups of $r_1$ and $r_2$ values and then calculate weighted averages assuming the $r_1$ and $r_2$ are selected independently random
from the empirical distribution of activities in our data.

We display results among different groups of activity in Fig.~\ref{sratios_group} and also in Table~S1 in the 
Supplementary Material. We see that high success ratios require relatively high activity; activities which we might consider typical, e.g.~between 30 
and 39 or 40 and 49 records in the transportation data (corresponding to 2-4 trips every day; recall that most trip results in two records in our 
dataset) and between 150 and 199 records in the phone data (one call or text per hour on average) only lead to $14\%$ -- $24\%$ success rates. Weighted 
average of estimated success rates of user groups with these activity (i.e.~with any random activity level in the other dataset) are in a similar 
range, while weighted average success ratio for the whole dataset (i.e.~choosing two trajectories from the two datasets randomly and assuming they 
would be a real match, i.e.~belong to the same person) is only around $8.1\%$. We also see in Fig.~\ref{sratios_group} that success ratios sharply increase as the number of 
records in the trajectories increase. Leaving out very inactive users from the averages (people with less than 10 transportation events and less than 
20 mobile communication records) increases the weighted average success ratio to $14.8\%$ already. 
Looking at users with very high activity in the communication dataset having over 1000 records and still focusing on typical transportation 
users whose weekly number of records is between 30 and 49, our estimation yields success ratios of over $90\%$. We note that only considering calls and text messages, these number of activities can be considered unrealistically high 
for typical people. Such frequencies could be achieved by using detailed CDRs including data communication and cell handover information, suggesting that achieving such success rates could be feasible with data that is currently already available to operators.

\begin{figure}
	\includegraphics{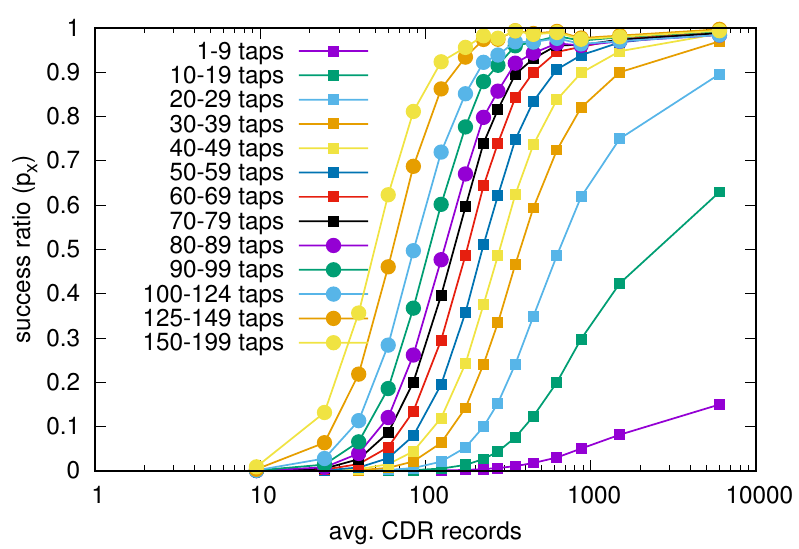}
	\caption{{\bf Estimated success ratios} for the one week long dataset for different activity level of users. The $x$-axis corresponds to activity
		in the communications dataset, while the different lines correspond to different level of activity in the transportation dataset.}
	\label{sratios_group}
\end{figure}

\begin{figure*}
\centering
	\includegraphics{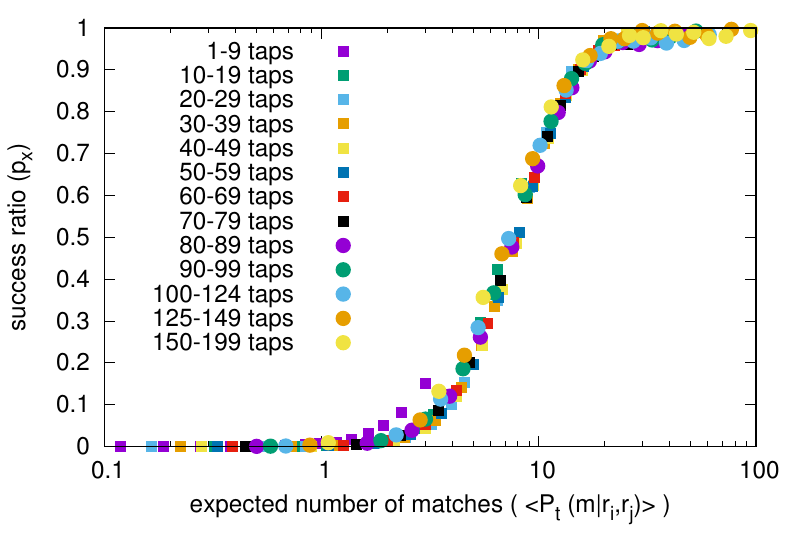}
	\includegraphics{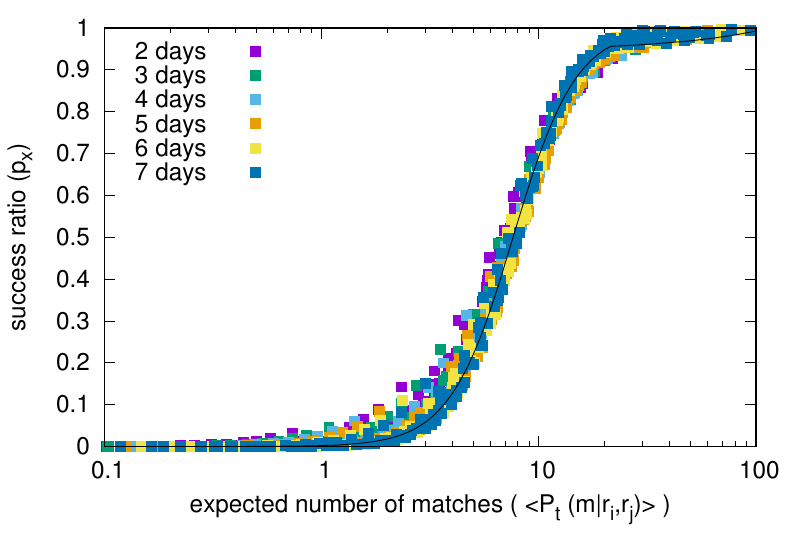}
	\caption{{\bf Estimated success ratios as a function of the expected number of matches.} The expected number of matches was estimated from the
		data, as the average number of temporal matches found between the different user groupd. Left: data is displayed for the whole week long 
		dataset, colors and shape of points correspond to different activity in the transportation dataset. We see that all of them follow the same 
		relation approximately. Right: data is displayed for the one week long dataset along with subsets of shorter time intervals and colors 
		correspond to different time intervals (e.g.~the dark blue points include all data from the left panel). Again, we see that all different 
		intervals follow approximately the same relation, allowing us to extrapolate to longer time intervals where $P_s$ is not known, only the 
		expected number of matches, $\bar{m} \equiv <P_t (m | r_i, r_j)>$ can be extrapolated. The black line displays a fitted analytical function
		we use for this as explained in the main text.}
	\label{sratios_tmatches_d17}
	\label{sratios_tmatches_all}
\end{figure*}

To better characterize how matchability depends on activities, in Fig.~\ref{sratios_tmatches_all} we display success ratios among all user pair
groups as the function of the expected number of temporal matches among those groups, i.e.~$<P_t (m | r_i, r_j)>$. We see that all different cases
follow the same relation, suggesting that the number of expected matches (or the number of real matches in case of looking at the representation of
a specific person in the data) is the variable determining matchability. Using this as a working hypothesis, we can extrapolate to longer time
intervals by employing the assumption that the dependence of success ratios on the expected number of matches does not change significantly with the
time interval considered. We test this by performing the same analysis for shorter sub-intervals of the one week data that is available for us, and
displaying the dependence of success ratios on the average number of temporal matches among different groups of users in Fig.~\ref{sratios_tmatches_d17}.
We see that for all cases with intervals ranging from $2$ to $7$ days, we have a very similar behavior, with somewhat larger variations for shorter
time intervals as it is expected as a consequence of more limited amount of available data. This relation can be fitted well with the analytical form
of $p_x (r_i,r_j) = \frac{1}{1 + A \bar{m}^{-b} }$, where $\bar{m} \equiv <P_t (m | r_i, r_j)>$ is the expected number of matches and the fitted
parameters are $b = 2.993$ and $A = 434.69$. For high values of $\bar{m}$, we use a more conservative estimate, given as a linear function: $p_x (r_i,
r_j) = 4.66 \cdot 10^{-4} \bar{m} + 0.946$ for $\bar{m} > 21.09$, so as not to overestimate success ratios in this area where points in the one week
long dataset are more scattered. We note that this choice does not alter the end results significantly. Furthermore, to show that the expected number
of false positive matches will not increase significantly as we increase the observation interval, in Fig.~\ref{spratios2}, we show the ratio of 
spatially consistent and total temporal matches among any two users in the dataset for these different time intervals. We see that this behavior is 
very similar for different time intervals as well, with a sharp decrease in spatially consistent trajectories as the number of matches increases, 
suggesting that as the expected number of matches among users increases, the number of false positive matches will continue to decrease. This supports
our extrapolation methodology by helping to establish that it will not overestimate success ratios.

We then perform the extrapolation by assuming that the expected number of temporal matches among the groups scales linearly with time (which
essentially corresponds to a convolution of the $P_t$ distributions for the longer time intervals), and interpolate the expected success ratios
as a function of the expected number of temporal matches using the previous simple functional form which fits the data well. Based on this, we
calculate similar measures as in the case of the original one week long dataset up to four weeks; we display the individual values among different
groups of users in Tables~S2, S3 and~S4 in the Supplementary Material, while summarize the results in Table~\ref{tab_extrapolating}.

We can make several observations based on these results which we can use to project the possibility of reidentification in several different scenarios
in terms of data collection methodology and data density. While readers are encouraged to look into Tables~S1--S4 in the Supplementary Material for
more insights, here we summarize the cases we find most relevant:

	\paragraph*{1) \emph{Matching transportation and CDR data.}} This is essentially the case of datasets we have at hand, and here we focus on regular
		transportation users (people with 30 -- 39 taps per week, which corresponds to 2 -- 3 trips taken per day). If we match this with a typical
		number of phone calls and messages (5 -- 10 per day, resulting in 30 -- 69 records per week), success ratios are generally low, $<1\%$ for the
		one week long dataset, and only increase to between $18\% to 43\%$ when considering a four week long period. Nevertheless, if phone
		activity increases to between 21 -- 28 records per day (150 -- 199 records per week, i.e.~one record per hour on average), success of matching
		is already $14\%$ for one week and reaches over $92\%$ for a four week data collection interval. We emphasize that about $34\%$ of all phone users in our dataset have at least 150 records per week, suggesting that such considerations are reasonable for a significant portion of the population. Furthermore, allowing even longer data collection intervals would result in identifying even less active users as well. In the case
		of people who we consider typical transportation users and moderate phone users (30 -- 39 records per week in the transportation dataset and
		50 -- 69 records per week in the mobile phone dataset), we estimate that after $11$ weeks, success rate would reach around $95\%$.
	\paragraph*{2) \emph{Matching transportation or similar dataset with detailed CDR or GPS traces.}} With the proliferation of smartphones and data connections,
		people generate much higher activity in the mobile network than it used to be the case. Even when not actively using a smartphone, apps running
		in the background periodically check for updates, generating data traffic which is logged by the network. Furthermore, many apps record
		location periodically (as reported by the phone based either on GPS or wireless signal) and report it to the app developers. Both cases allow
		a much higher quality reconstruction of the people's trajectory during the day. Using a conservative estimate, we can expect this case to
		correspond to data collection with at least one point per hour, yielding a similar result as for people considered ``active'' phone users in the previous case, i.e.~an expected success of matching with typical transportation
		users of at least $14\%$ for one week and over $92\%$ for a four weeks. Making similar estimates with doubled data collection rate (once per
		half an hour, or about 300 -- 399 records per week), we have an expected success rate of $46\%$ for only one week already, almost $90\%$ for
		two weeks and over $95\%$ for three or four weeks. We note that these are still conservative estimates, as both network operators and app developers can easily detect users' movements and implement adaptive data collection, allowing them to reconstruct trajectories in good quality with less data points.
		We further note that beside transportation data, credit and debit card usage can generate similar amount
		of records in developed countries where a major portion of payments is made electronically, as well as geo-tagged social media posts of people
		actively maintaining a presence on microblogging services.
	\paragraph*{3) \emph{Matching two datasets with increased density.}} Based on the previous discussion that data traffic and smartphone apps running in the background can easily generate at least one
		record per hour (i.e.~between 150 and 199 records per week), we can argue that having access to two such datasets should be considered possible.
		Looking
		at results between two groups both having between 150 and 199 records per week, we see that the expected success ratio is already $95.6\%$ even
		for only one week of data collection. Assuming somewhat fewer records (between 100 and 124), success ratio for one week is $72\%$ and reaches
		almost $95\%$ already after two weeks, establishing that the data collection procedures easily implementable for any smartphone app developer
		already generate data which allows reidentification only after a very short data collection interval.

We emphasize that the main basis for matchability is that the probability of a temporal match with \emph{one} user ($P_t$, which we then consider to
describe the distribution of real matches) should be higher than the probability of a random spatial match occurring with \emph{any} of the $\sim 2.8$
million users in the other dataset (described by the $\gamma$ matchability parameter estimated from $P_s$). Whether this holds true depends on the
statistics of the dataset and estimating these correctly requires that the mobility traces of all of the affected population be present so as to be
able to obtain a realistic estimate of false positive spatial matches. In our case, Figs.~\ref{spratios} and~\ref{matchescmp} illustrate that the
ratio of probabilities of spatial and temporal matches decreases to small values quickly and that these probabilities are indeed comparable.

\begin{table*}
\centering
\small
	\begin{tabular}{r|r|r|r|r}
		 & \parbox{2cm}{success ratio (all groups)} & \parbox{3cm}{success ratio without inactive users} 
			& \parbox{4cm}{success ratio for transportation users with 30--49 records}
			& \parbox{4cm}{success ratio for transportation users with 30--49 records without inactive CDR users} \\ \hline
		 1 week  & $0.0805$ & $0.1484$ & $0.1677$ & $0.2181$ \\
		 2 weeks & $0.1657$ & $0.3062$ & $0.3718$ & $0.4882$ \\
		 3 weeks & $0.2225$ & $0.41$   & $0.4827$ & $0.6348$ \\
		 4 weeks & $0.2588$ & $0.4766$ & $0.551$  & $0.7248$ 
	\end{tabular}
	\caption{{\bf Average expected success ratios} for the study data and extrapolated to longer intervals. The left column shows results from
		any possible combination of activity pairings (i.e.~assuming that the number records corresponding to a person's activities in the two
		datasets are randomly selected among all possible users activity levels in the two dataset). The right column shows averages with leaving
		out users with very low activities (less than 10 taps in the case of the transportation dataset and less than 20 records in the case of
		the mobile communication data) and also users with unrealistically high activities (125 or more taps in the transportation dataset or 2000
		or more records in the mobile phone dataset).
		}
	\label{tab_extrapolating}
\end{table*}

\section{Discussion}
\label{sec_conclusion}

In this paper we considered the problem of \emph{matching} mobility datasets on a realistic scale in an urban setting. We developed methodology
for handling the problem of comparing users' trajectories on the scale of several hundred million records of several million users and applied our
solution for a dataset of one week of mobility traces of mobile communications and transportation logs from Singapore. We presented an empirical
framework for calculating the estimated success ratio of matching users based on co-locations. Our results suggest that trajectories of people who have typical transportation usage patterns and are relatively active phone users could be matched based on a few weeks of data, while matching two datasets where data is collected more
regularly (i.e.~having one record per hour on average, which is easily achieved by network operators logging data communications or smartphone
applications regularly querying device location) can be easily possible based on only one week of data collection. As the trend of collecting spatial
traces of people continues with many service providers, we expect the possibility of matching people in anonymized datasets based on their trajectories
to become increasingly easy in the near future.

Comparing our work to previous studies, we believe that this is the first study which tried to estimate the chances of successfully matching
trajectories involving two large-scale dataset (on the order of millions of people in each) from realistic sources. We believe that using data
that covers a significant portion of the population is important for this kind of work as data density is a main contributing factor to the
success of matching, as with larger datasets we can expect significantly more false positive matches. In our analysis, the probability of
finding a certain number of false positive matches is represented by the probability $P_{s}$ for one user pair and with the $\gamma$ matchability
parameter for the whole dataset. While we saw that $P_{s}$ is a quickly decreasing function
of the required number of matches, $\gamma$ depends exponentially on the number of users $n_{C}$ and thus the density of data, meaning that
the probability of finding false positive matches can drastically differ between datasets of different density. Furthermore, most of the methods
employed for calculating matches in previous work (e.g. in Refs.~\cite{Gambs2014,Riederer2016}) employ a metric which needs to be calculated for
each possible pair of users, thus being prohibitably computationally intensive in the case of realistic data densities as they result in
$O(n_{T} \times n_{C})$ complexity. For the aforementioned reasons, we believe that scaling down our methods for the smaller data sizes used in these works is not
expected to give realistic estimates. This way, a more direct comparison of our work with these papers is not easy to achieve.

The work of Cao~et~al.~\cite{Cao2016} presents a computationally efficient implementation, but disregards information about
temporal matches, thus actually considering a much simpler version of the problem than what we focus on. The work of Cecaj~et~al.~\cite{Cecaj2015}
is more similar to ours, but is also applied to a more limited data source, limiting the potential statistical analysis. We believe that our work
is most similar to that of Bas\i{}k~et~al.~\cite{Basik2017} who describe a very similar matching procedure augmented by an efficient spatial
pre-filtering to reduce complexity. In our case, we believe such pre-filtering would not be applicable as our data comes from a compact but
densely populated area, thus all users' trajectories would span the same units when constructing a coarse spatial index for such pre-filtering.
The complexity of the main matching procedure is very similar; while they compute a different metric to evaluate potential
candidates, enumerating matches is based on similar computational steps. There is a significant difference however in the primary focus of our current study and Ref.~\cite{Basik2017}. The authors in Ref.~\cite{Basik2017} primarily focus on evaluating the matching algorithm in terms of computational performance on real data, while they only test the quality of matching on synthetic data. On the contrary, our main focus is estimating the expected success of merging two real-world large scale datasets based on the statistical analysis of the matches found among them. To this end, we used a simpler definition of matching which allows the derivation of probabilities for finding false positive matches which we believe is highly determinant in matching perfomance. The main difference is that the authors of Ref.~\cite{Basik2017} not only count the number of matches, but also the number of distinct locations such matches occur at. In our framework, that would mean an additional random variable for the $P_s$ probability distribution, which would make the empirical estimation much more unreliable. This way, we believe our results represent only a lower bound on the success rates which could be improved upon if training on ground truth data becomes possible. Thus our results are not directly comparable to the matching quality quality metrics in Ref.~\cite{Basik2017}, since those are influenced by the procedure used to generate the synthetic dataset.

Future work can extend the matching procedure employed, i.e.~instead of just selecting the candidate with the highest number of matches, a more
sophisticated approach could take into account the uneven nature of urban movements and calculate for each matching pair of points a weight
representing the importance of it (e.g.~a match at a crowded subway station could be considered less important than a match at a remote bus stop) or take into account the number of distinct places trajectories are matched similar to Ref.~\cite{Basik2017}. We believe such refinements would require at least some ground truth data such that quality of matches can be readily evaluated and optimized.
Our matchability estimate could be adapted to a such scenario as long as statistics of matching points can be calculated similarly to the $P_t$ and $P_s$ distributions, i.e.~assuming independence and identical distributions for large subsets of users allowing empirical estimate of such distributions. Using a dataset with available ground truth information could also be utilized to calculate measures of individual
matchability and establish a connection with the entropy and predictability measures defined in previous work of Song et al.~\cite{Song2010}.

We believe that the possibility of matching mobility traces can open up new potential for understanding urban human mobility and providing better
services for urban residents. Utilizing this while also providing adequate guarantees of privacy of the affected individuals should be in the focus
of future interdisciplinary research including urban planning, algorithmic, security and legal perspectives.

\section*{Acknowledgements}

\small
The authors thank Ali Farzanehfar and Arnaud Tournier for reading the manuscript and providing valuable comments and suggestions.

The authors thank all sponsors and partners of the MIT Senseable City Laboratory including Allianz, the Amsterdam Institute for Advanced Metropolitan Solutions, the Fraunhofer Institute, Kuwait-MIT Center for Natural Resources and the Environment, Singapore-MIT Alliance for Research and Technology (SMART) and all the members of the Consortium.

\clearpage



\section*{Supplementary material}

## 1 week

| Activity | \multicolumn{13}{c}{transportation user activity (number of taps per week)} | average | limited average (users with between 10 – 124 taps / week) |
|---|---|---|---|---|---|---|---|---|---|---|---|---|---|---|---|
| | 1–9 (1,047,508) | 10 – 19 (746,005) | 20 – 29 (618,935) | 30 – 39 (412,302) | 40 – 49 (259,683) | 50 – 59 (147,204) | 60 – 69 (68,053) | 70 – 79 (29,359) | 80 – 89 (11,889) | 90 – 99 (4,425) | 100 – 124 (2,730) | 125 –149 (380) | 150 – 199 (122) | | |
| 0 – 19 (687435) | $1.98 \cdot 10^{-6}$ (0.0371) | $2.19 \cdot 10^{-6}$ (0.0999) | $4.02 \cdot 10^{-6}$ (0.164) | $1.21 \cdot 10^{-5}$ (0.225) | $2.65 \cdot 10^{-5}$ (0.279) | $5.44 \cdot 10^{-5}$ (0.332) | $1.16 \cdot 10^{-4}$ (0.388) | $2.16 \cdot 10^{-4}$ (0.445) | $3.41 \cdot 10^{-4}$ (0.502) | $6.49 \cdot 10^{-4}$ (0.581) | $1.37 \cdot 10^{-3}$ (0.683) | $3.52 \cdot 10^{-3}$ (0.882) | $9.13 \cdot 10^{-3}$ (1.07) | $1.6 \cdot 10^{-5}$ | $2.125 \cdot 10^{-5}$ |
| 20 – 29 (155136) | $2.18 \cdot 10^{-5}$ (0.118) | $3.81 \cdot 10^{-5}$ (0.321) | $8.90 \cdot 10^{-5}$ (0.528) | $2.83 \cdot 10^{-4}$ (0.724) | $6.31 \cdot 10^{-4}$ (0.902) | $1.31 \cdot 10^{-3}$ (1.07) | $2.74 \cdot 10^{-3}$ (1.26) | $5.02 \cdot 10^{-3}$ (1.44) | $7.80 \cdot 10^{-3}$ (1.62) | 0.0142 (1.87) | 0.0277 (2.2) | 0.0636 (2.84) | 0.1318 (3.47) | $3.5 \cdot 10^{-4}$ | $4.872 \cdot 10^{-4}$ |
| 30 – 49 (244241) | $6.60 \cdot 10^{-5}$ (0.188) | $1.67 \cdot 10^{-4}$ (0.512) | $4.86 \cdot 10^{-4}$ (0.846) | $1.66 \cdot 10^{-3}$ (1.16) | $3.74 \cdot 10^{-3}$ (1.45) | $7.63 \cdot 10^{-3}$ (1.73) | 0.0152 (2.02) | 0.0263 (2.31) | 0.0390 (2.6) | 0.0656 (3) | 0.1133 (3.53) | 0.2182 (4.54) | 0.3562 (5.54) | 0.0019 | $2.641 \cdot 10^{-3}$ |
| 50 – 69 (202655) | $1.68 \cdot 10^{-4}$ (0.276) | $5.72 \cdot 10^{-4}$ (0.759) | $1.96 \cdot 10^{-3}$ (1.26) | $6.74 \cdot 10^{-3}$ (1.73) | 0.0148 (2.16) | 0.0287 (2.58) | 0.0532 (3.01) | 0.0861 (3.45) | 0.1204 (3.88) | 0.1859 (4.48) | 0.2839 (5.25) | 0.4606 (6.76) | 0.6232 (8.24) | 0.0066 | $9.417 \cdot 10^{-3}$ |
| 70 – 99 (256755) | $3.79 \cdot 10^{-4}$ (0.379) | $1.67 \cdot 10^{-3}$ (1.05) | $6.39 \cdot 10^{-3}$ (1.75) | 0.0212 (2.4) | 0.0438 (3) | 0.0793 (3.57) | 0.1346 (4.18) | 0.2004 (4.79) | 0.2614 (5.38) | 0.3671 (6.2) | 0.4969 (7.26) | 0.6876 (9.36) | 0.8114 (11.4) | 0.0176 | 0.0253 |
| 100 – 149 (330530) | $9.49 \cdot 10^{-4}$ (0.524) | $5.25 \cdot 10^{-3}$ (1.45) | 0.0209 (2.43) | 0.0632 (3.34) | 0.1193 (4.19) | 0.1948 (4.98) | 0.2950 (5.83) | 0.3960 (6.68) | 0.4765 (7.5) | 0.6017 (8.65) | 0.7198 (10.1) | 0.8623 (13.1) | 0.9237 (15.9) | 0.0436 | 0.0628 |
| 150 – 199 (237813) | $2.12 \cdot 10^{-3}$ (0.684) | 0.0135 (1.9) | 0.0527 (3.19) | 0.1419 (4.39) | 0.2420 (5.5) | 0.3572 (6.55) | 0.4866 (7.66) | 0.5957 (8.78) | 0.6703 (9.87) | 0.7767 (11.4) | 0.8516 (13.3) | 0.9340 (17.2) | 0.9560 (21) | 0.0847 | 0.1221 |
| 200 – 249 (170137) | $3.90 \cdot 10^{-3}$ (0.846) | 0.0268 (2.36) | 0.0996 (3.95) | 0.2401 (5.44) | 0.3750 (6.82) | 0.5105 (8.13) | 0.6432 (9.51) | 0.7398 (10.9) | 0.7981 (12.3) | 0.8789 (14.1) | 0.9222 (16.6) | 0.9741 (21.5) | 0.9822 (26.2) | 0.1311 | 0.1889 |
| 250 – 299 (123475) | $6.13 \cdot 10^{-3}$ (0.974) | 0.0436 (2.72) | 0.1528 (4.56) | 0.3341 (6.28) | 0.4855 (7.88) | 0.6212 (9.39) | 0.7397 (11) | 0.8158 (12.6) | 0.8572 (14.2) | 0.9156 (16.3) | 0.9389 (19.2) | 0.9743 (24.8) | 0.9769 (30.2) | 0.1733 | 0.2493 |
| 300 – 399 (157376) | 0.0105 (1.17) | 0.0754 (3.26) | 0.2405 (5.47) | 0.4656 (7.54) | 0.6237 (9.47) | 0.7477 (11.3) | 0.8426 (13.2) | 0.8953 (15.2) | 0.9206 (17.1) | 0.9598 (19.7) | 0.9687 (23.1) | 0.9938 (29.9) | 0.9934 (36.4) | 0.2335 | 0.3349 |
| 400 – 499 (89358) | 0.0175 (1.37) | 0.1232 (3.82) | 0.3485 (6.42) | 0.5935 (8.85) | 0.7363 (11.1) | 0.8343 (13.3) | 0.8999 (15.6) | 0.9307 (17.9) | 0.9433 (20.1) | 0.9683 (23.2) | 0.9684 (27.2) | 0.9874 (35.2) | 0.9856 (43) | 0.2962 | 0.4230 |
| 500 – 749 (102698) | 0.0303 (1.64) | 0.1998 (4.58) | 0.4865 (7.7) | 0.7245 (10.6) | 0.8383 (13.4) | 0.9065 (16) | 0.9460 (18.8) | 0.9609 (21.6) | 0.9651 (24.3) | 0.9814 (28) | 0.9766 (32.8) | 0.9922 (42.3) | 0.9897 (51.6) | 0.3713 | 0.5264 |
| 750 – 999 (39523) | 0.0502 (1.93) | 0.2975 (5.38) | 0.6186 (9.07) | 0.8199 (12.6) | 0.8984 (15.9) | 0.9384 (19) | 0.9574 (22.3) | 0.9614 (25.6) | 0.9604 (28.9) | 0.9714 (33.2) | 0.9638 (38.7) | 0.9774 (49.6) | 0.9747 (60.7) | 0.4417 | 0.6199 |
| 1000 – 1999 (38714) | 0.0815 (2.32) | 0.4234 (6.45) | 0.7499 (10.9) | 0.9002 (15.1) | 0.9475 (19.2) | 0.9673 (23.1) | 0.9745 (27.1) | 0.9731 (31.2) | 0.9697 (35.2) | 0.9786 (40.2) | 0.9701 (46.6) | 0.9829 (59.4) | 0.9801 (72.7) | 0.5193 | 0.7185 |
| 2000 – 9999 (8791) | 0.1497 (3) | 0.6295 (8.34) | 0.8956 (14.2) | 0.9699 (19.8) | 0.9872 (25.3) | 0.9914 (30.6) | 0.9918 (35.9) | 0.9881 (41.2) | 0.9838 (46.5) | 0.9925 (52.9) | 0.9838 (60.9) | 0.9967 (76.9) | 0.9939 (94.4) | 0.6267 | 0.8439 |
| average | 0.0053 | 0.0325 | 0.0836 | 0.1478 | 0.1993 | 0.2495 | 0.3016 | 0.3454 | 0.3774 | 0.4289 | 0.4763 | 0.5486 | 0.5976 | 0.0805 | |
| limited average (users with between 20 – 1999 CDR records / week) | 0.0381 | 0.2383 | 0.4391 | 0.6125 | 0.7110 | 0.7812 | 0.8355 | 0.8704 | 0.8874 | 0.9090 | 0.9241 | 0.9383 | 0.9435 | | 0.1484 |

(Row label on left side of table: CDR user activity (number of records per week))

Table S1: **Success ratios for different groups of users by activity**, estimated for the original, one week long dataset. Numbers in the table correspond to the success ratios in that group. Numbers in parentheses below the success ratios are the expected number of matches in that group. Numbers in parentheses below the user activities are the number of people in that group. The averages are calculated by weighting by the number of users. The last row and column displays averages calculated with leaving out casual users with very low activity (for whom this matching procedure is not expected to be possible), and users with unrealistically high activiy.

Supplementary information for *Towards matching user mobility traces in large-scale datasets*

## 2 weeks

| Activity | \multicolumn{13}{c}{transportation user activity (number of taps per week)} | average | limited average (users with between 10 – 124 taps / week) |
|---|---|---|---|---|---|---|---|---|---|---|---|---|---|---|---|
| | 1–9 (1,047,508) | 10 – 19 (746,005) | 20 – 29 (618,935) | 30 – 39 (412,302) | 40 – 49 (259,683) | 50 – 59 (147,204) | 60 – 69 (68,053) | 70 – 79 (29,359) | 80 – 89 (11,889) | 90 – 99 (4,425) | 100 – 124 (2,730) | 125 –149 (380) | 150 – 199 (122) | | |
| 0 – 19 (687435) | $9.54 \cdot 10^{-7}$ (0.0741) | $1.85 \cdot 10^{-5}$ (0.2) | $8.16 \cdot 10^{-5}$ (0.328) | $2.09 \cdot 10^{-4}$ (0.449) | $4.01 \cdot 10^{-4}$ (0.558) | $6.75 \cdot 10^{-4}$ (0.664) | $1.08 \cdot 10^{-3}$ (0.777) | $1.62 \cdot 10^{-3}$ (0.89) | $2.32 \cdot 10^{-3}$ (1) | $3.59 \cdot 10^{-3}$ (1.16) | $5.82 \cdot 10^{-3}$ (1.37) | 0.0124 (1.76) | 0.0221 (2.15) | $1.62 \cdot 10^{-4}$ | $2.32 \cdot 10^{-4}$ |
| 20 – 29 (155136) | $3.07 \cdot 10^{-5}$ (0.236) | $6.09 \cdot 10^{-4}$ (0.642) | $2.70 \cdot 10^{-3}$ (1.06) | $6.93 \cdot 10^{-3}$ (1.45) | 0.0133 (1.8) | 0.0221 (2.15) | 0.0350 (2.51) | 0.0517 (2.88) | 0.0721 (3.24) | 0.1072 (3.75) | 0.1628 (4.4) | 0.2943 (5.68) | 0.4305 (6.93) | $5.25 \cdot 10^{-3}$ | $7.55 \cdot 10^{-3}$ |
| 30 – 49 (244241) | $1.22 \cdot 10^{-4}$ (0.375) | $2.46 \cdot 10^{-3}$ (1.02) | 0.0110 (1.69) | 0.0279 (2.32) | 0.0527 (2.9) | 0.0858 (3.45) | 0.1304 (4.04) | 0.1841 (4.63) | 0.2428 (5.2) | 0.3303 (6.01) | 0.4432 (7.05) | 0.6296 (9.08) | 0.7546 (11.1) | 0.0199 | 0.0288 |
| 50 – 69 (202655) | $3.88 \cdot 10^{-4}$ (0.552) | $7.95 \cdot 10^{-3}$ (1.52) | 0.0354 (2.52) | 0.0867 (3.47) | 0.1560 (4.33) | 0.2374 (5.15) | 0.3321 (6.03) | 0.4277 (6.9) | 0.5149 (7.76) | 0.6193 (8.95) | 0.7236 (10.5) | 0.8482 (13.5) | 0.9100 (16.5) | 0.0555 | 0.0804 |
| 70 – 99 (256755) | $1.00 \cdot 10^{-3}$ (0.757) | 0.0206 (2.09) | 0.0886 (3.49) | 0.2014 (4.8) | 0.3299 (6) | 0.4532 (7.15) | 0.5695 (8.35) | 0.6652 (9.57) | 0.7381 (10.8) | 0.8116 (12.4) | 0.8739 (14.5) | 0.9367 (18.7) | 0.9568 (22.8) | 0.1135 | 0.1646 |
| 100 – 149 (330530) | $2.64 \cdot 10^{-3}$ (1.05) | 0.0531 (2.91) | 0.2074 (4.86) | 0.4047 (6.69) | 0.5711 (8.37) | 0.6917 (9.97) | 0.7818 (11.7) | 0.8434 (13.4) | 0.8842 (15) | 0.9211 (17.3) | 0.9495 (20.3) | 0.9584 (26.1) | 0.9611 (31.8) | 0.2041 | 0.2956 |
| 150 – 199 (237813) | $5.83 \cdot 10^{-3}$ (1.37) | 0.1116 (3.81) | 0.3711 (6.38) | 0.6057 (8.78) | 0.7510 (11) | 0.8358 (13.1) | 0.8905 (15.3) | 0.9244 (17.6) | 0.9456 (19.7) | 0.9568 (22.8) | 0.9587 (26.7) | 0.9623 (34.4) | 0.9658 (42) | 0.2966 | 0.4288 |
| 200 – 249 (170137) | 0.0110 (1.69) | 0.1927 (4.72) | 0.5286 (7.91) | 0.7449 (10.9) | 0.8517 (13.6) | 0.9065 (16.3) | 0.9394 (19) | 0.9564 (21.8) | 0.9576 (24.5) | 0.9594 (28.3) | 0.9617 (33.2) | 0.9662 (42.9) | 0.9706 (52.3) | 0.3748 | 0.5403 |
| 250 – 299 (123475) | 0.0167 (1.95) | 0.2673 (5.43) | 0.6321 (9.12) | 0.8176 (12.6) | 0.8983 (15.8) | 0.9373 (18.8) | 0.9565 (22) | 0.9580 (25.2) | 0.9594 (28.4) | 0.9614 (32.7) | 0.9641 (38.4) | 0.9693 (49.6) | 0.9744 (60.5) | 0.4266 | 0.6131 |
| 300 – 399 (157376) | 0.0283 (2.33) | 0.3860 (6.52) | 0.7480 (10.9) | 0.8858 (15.1) | 0.9388 (18.9) | 0.9567 (22.6) | 0.9585 (26.5) | 0.9604 (30.3) | 0.9621 (34.2) | 0.9646 (39.4) | 0.9678 (46.2) | 0.9741 (59.7) | 0.9802 (72.9) | 0.4906 | 0.7009 |
| 400 – 499 (89358) | 0.0448 (2.74) | 0.5032 (7.64) | 0.8269 (12.8) | 0.9260 (17.7) | 0.9566 (22.2) | 0.9586 (26.6) | 0.9607 (31.1) | 0.9629 (35.7) | 0.9650 (40.2) | 0.9678 (46.4) | 0.9716 (54.5) | 0.9790 (70.4) | 0.9863 (85.9) | 0.5429 | 0.7696 |
| 500 – 749 (102698) | 0.0746 (3.28) | 0.6350 (9.16) | 0.8920 (15.4) | 0.9561 (21.3) | 0.9587 (26.8) | 0.9612 (32.1) | 0.9637 (37.6) | 0.9663 (43.2) | 0.9689 (48.7) | 0.9723 (56) | 0.9768 (65.6) | 0.9856 (84.5) | 0.9943 (103) | 0.5977 | 0.8358 |
| 750 – 999 (39523) | 0.1164 (3.87) | 0.7385 (10.8) | 0.9308 (18.1) | 0.9579 (25.1) | 0.9610 (31.7) | 0.9640 (38.1) | 0.9670 (44.7) | 0.9701 (51.3) | 0.9732 (57.8) | 0.9771 (66.4) | 0.9823 (77.4) | 0.9925 (99.3) | 1.0000 (121) | 0.6417 | 0.8808 |
| 1000 – 1999 (38714) | 0.1862 (4.65) | 0.8291 (12.9) | 0.9564 (21.8) | 0.9603 (30.3) | 0.9641 (38.4) | 0.9678 (46.2) | 0.9715 (54.3) | 0.9753 (62.3) | 0.9790 (70.3) | 0.9837 (80.4) | 0.9897 (93.3) | 1.0000 (119) | 1.0000 (145) | 0.6893 | 0.9183 |
| 2000 – 9999 (8791) | 0.3289 (6) | 0.9128 (16.7) | 0.9594 (28.4) | 0.9647 (39.7) | 0.9698 (50.5) | 0.9747 (61.1) | 0.9797 (71.8) | 0.9846 (82.4) | 0.9895 (93) | 0.9955 (106) | 1.0000 (122) | 1.0000 (154) | 1.0000 (189) | 0.7547 | 0.9486 |
| average | 0.0131 | 0.1257 | 0.2543 | 0.3469 | 0.4114 | 0.4590 | 0.4991 | 0.5316 | 0.5578 | 0.5875 | 0.6181 | 0.6605 | 0.6893 | 0.1921 | |
| limited average (users with between 20 – 1999 CDR records / week) | 0.0161 | 0.1628 | 0.3327 | 0.4552 | 0.5406 | 0.6036 | 0.6565 | 0.6994 | 0.7337 | 0.7726 | 0.8125 | 0.8664 | 0.9015 | | 0.3581 |

Row labels (leftmost column): CDR user activity (number of records per week)

Table S2: **Success ratios for different groups of users by activity**, extrapolated for two weeks. Numbers in the table correspond to the success ratios in that group. Numbers in parentheses below the success ratios are the expected number of matches in that group. Numbers in parentheses below the user activities are the number of people in that group. The averages are calculated by weighting by the number of users. The last row and column displays averages calculated with leaving out casual users with very low activity (for whom this matching procedure is not expected to be possible), and users with unrealistically high activiy.

## 3 weeks

| | | transportation user activity (number of taps per week) | | | | | | | | | | | | | | limited average |
|---|---|---|---|---|---|---|---|---|---|---|---|---|---|---|---|---|
| Activity | 1–9 (1,047,508) | 10–19 (746,005) | 20–29 (618,935) | 30–39 (412,302) | 40–49 (259,683) | 50–59 (147,204) | 60–69 (68,053) | 70–79 (29,359) | 80–89 (11,889) | 90–99 (4,425) | 100–124 (2,730) | 125–149 (380) | 150–199 (122) | average | (users with between 10–124 taps / week) |
|---|---|---|---|---|---|---|---|---|---|---|---|---|---|---|---|
| 0–19 (687435) | $3.21 \cdot 10^{-6}$ (0.111) | $6.23 \cdot 10^{-5}$ (0.3) | $2.75 \cdot 10^{-4}$ (0.492) | $7.04 \cdot 10^{-4}$ (0.674) | $1.35 \cdot 10^{-3}$ (0.837) | $2.27 \cdot 10^{-3}$ (0.996) | $3.62 \cdot 10^{-3}$ (1.16) | $5.44 \cdot 10^{-3}$ (1.34) | $7.75 \cdot 10^{-3}$ (1.5) | 0.0120 (1.74) | 0.0193 (2.05) | 0.0407 (2.65) | 0.0708 (3.22) | $5.44 \cdot 10^{-4}$ | $7.80 \cdot 10^{-4}$ |
| 20–29 (155136) | $1.03 \cdot 10^{-4}$ (0.355) | $2.05 \cdot 10^{-3}$ (0.963) | $9.05 \cdot 10^{-3}$ (1.58) | 0.0230 (2.17) | 0.0433 (2.71) | 0.0708 (3.22) | 0.1087 (3.77) | 0.1551 (4.32) | 0.2072 (4.86) | 0.2878 (5.62) | 0.3957 (6.61) | 0.5840 (8.52) | 0.7179 (10.4) | 0.0166 | 0.0239 |
| 30–49 (244241) | $4.11 \cdot 10^{-4}$ (0.563) | $8.24 \cdot 10^{-3}$ (1.54) | 0.0361 (2.54) | 0.0882 (3.49) | 0.1578 (4.35) | 0.2400 (5.18) | 0.3354 (6.06) | 0.4316 (6.94) | 0.5191 (7.81) | 0.6241 (9.01) | 0.7282 (10.6) | 0.8512 (13.6) | 0.9119 (16.6) | 0.0563 | 0.0815 |
| 50–69 (202655) | $1.30 \cdot 10^{-3}$ (0.828) | 0.0263 (2.28) | 0.1099 (3.78) | 0.2422 (5.2) | 0.3835 (6.49) | 0.5117 (7.73) | 0.6260 (9.04) | 0.7156 (10.4) | 0.7813 (11.6) | 0.8456 (13.4) | 0.8981 (15.7) | 0.9495 (20.3) | 0.9577 (24.7) | 0.1324 | 0.1919 |
| 70–99 (256755) | $3.36 \cdot 10^{-3}$ (1.14) | 0.0660 (3.14) | 0.2466 (5.24) | 0.4591 (7.2) | 0.6236 (9.01) | 0.7361 (10.7) | 0.8166 (12.5) | 0.8699 (14.4) | 0.9046 (16.1) | 0.9355 (18.6) | 0.9564 (21.8) | 0.9593 (28.1) | 0.9622 (34.2) | 0.2282 | 0.3304 |
| 100–149 (330530) | $8.82 \cdot 10^{-3}$ (1.57) | 0.1587 (4.36) | 0.4683 (7.29) | 0.6959 (10) | 0.8176 (12.6) | 0.8831 (15) | 0.9234 (17.5) | 0.9477 (20) | 0.9567 (22.5) | 0.9583 (25.9) | 0.9604 (30.4) | 0.9645 (39.2) | 0.9685 (47.8) | 0.3452 | 0.4983 |
| 150–199 (237813) | 0.0194 (2.05) | 0.2972 (5.71) | 0.6652 (9.57) | 0.8379 (13.2) | 0.9103 (16.5) | 0.9448 (19.7) | 0.9569 (23) | 0.9585 (26.3) | 0.9600 (29.6) | 0.9621 (34.1) | 0.9649 (40) | 0.9703 (51.7) | 0.9756 (63) | 0.4440 | 0.6372 |
| 200–249 (170137) | 0.0361 (2.54) | 0.4455 (7.07) | 0.7906 (11.9) | 0.9076 (16.3) | 0.9508 (20.5) | 0.9576 (24.4) | 0.9595 (28.5) | 0.9615 (32.7) | 0.9634 (36.8) | 0.9660 (42.4) | 0.9694 (49.8) | 0.9762 (64.4) | 0.9828 (78.5) | 0.5178 | 0.7370 |
| 250–299 (123475) | 0.0540 (2.92) | 0.5512 (8.15) | 0.8526 (13.7) | 0.9378 (18.8) | 0.9572 (23.6) | 0.9593 (28.2) | 0.9616 (33) | 0.9638 (37.8) | 0.9660 (42.5) | 0.9691 (49) | 0.9730 (57.6) | 0.9809 (74.4) | 0.9885 (90.7) | 0.5628 | 0.7944 |
| 300–399 (157376) | 0.0892 (3.5) | 0.6791 (9.77) | 0.9090 (16.4) | 0.9568 (22.6) | 0.9595 (28.4) | 0.9620 (33.9) | 0.9647 (39.7) | 0.9674 (45.5) | 0.9701 (51.3) | 0.9737 (59.1) | 0.9785 (69.4) | 0.9880 (89.6) | 0.9972 (109) | 0.6155 | 0.8550 |
| 400–499 (89358) | 0.1364 (4.11) | 0.7732 (11.5) | 0.9415 (19.2) | 0.9586 (26.6) | 0.9618 (33.4) | 0.9648 (39.9) | 0.9680 (46.7) | 0.9712 (53.6) | 0.9743 (60.4) | 0.9786 (69.6) | 0.9843 (81.7) | 0.9954 (106) | 1.0000 (129) | 0.6578 | 0.8952 |
| 500–749 (102698) | 0.2134 (4.92) | 0.8541 (13.7) | 0.9570 (23.1) | 0.9611 (31.9) | 0.9650 (40.2) | 0.9686 (48.1) | 0.9725 (56.4) | 0.9764 (64.7) | 0.9802 (73) | 0.9854 (84) | 0.9921 (98.4) | 1.0000 (127) | 1.0000 (155) | 0.7037 | 0.9269 |
| 750–999 (39523) | 0.3073 (5.8) | 0.9048 (16.1) | 0.9589 (27.2) | 0.9638 (37.7) | 0.9684 (47.6) | 0.9728 (57.1) | 0.9774 (67) | 0.9821 (76.9) | 0.9866 (86.8) | 0.9926 (99.5) | 1.0000 (116) | 1.0000 (149) | 1.0000 (182) | 0.7457 | 0.9453 |
| 1000–1999 (38714) | 0.4351 (6.97) | 0.9423 (19.3) | 0.9614 (32.6) | 0.9674 (45.4) | 0.9730 (57.5) | 0.9785 (69.3) | 0.9842 (81.4) | 0.9898 (93.5) | 0.9954 (105) | 1.0000 (121) | 1.0000 (140) | 1.0000 (178) | 1.0000 (218) | 0.7958 | 0.9600 |
| 2000–9999 (8791) | 0.6226 (8.99) | 0.9579 (25) | 0.9660 (42.6) | 0.9739 (59.5) | 0.9815 (75.8) | 0.9889 (91.7) | 0.9964 (108) | 1.0000 (124) | 1.0000 (139) | 1.0000 (159) | 1.0000 (183) | 1.0000 (231) | 1.0000 (283) | 0.8611 | 0.9696 |
| average | 0.0366 | 0.2236 | 0.3719 | 0.4609 | 0.5172 | 0.5568 | 0.5893 | 0.6154 | 0.6360 | 0.6591 | 0.6830 | 0.7158 | 0.7387 | 0.2725 | |
| limited average (users with between 20–1999 CDR records / week) | 0.0459 | 0.2921 | 0.4883 | 0.6060 | 0.6804 | 0.7325 | 0.7750 | 0.8090 | 0.8356 | 0.8648 | 0.8941 | 0.9306 | 0.9513 | | 0.4987 |

CDR user activity (number of records per week)

Table S3: **Success ratios for different groups of users by activity**, extrapolated for three weeks. Numbers in the table correspond to the success ratios in that group. Numbers in parentheses below the success ratios are the expected number of matches in that group. Numbers in parentheses below the user activities are the number of people in that group. The averages are calculated by weighting by the number of users. The last row and column displays averages calculated with leaving out casual users with very low activity (for whom this matching procedure is not expected to be possible), and users with unrealistically high activiy.

## 4 weeks

| | Activity | \multicolumn{13}{c}{transportation user activity (number of taps per week)} | average | limited average (users with between 10 – 124 taps / week) |
|---|---|---|---|---|---|---|---|---|---|---|---|---|---|---|---|---|
| | | 1–9 (1,047,508) | 10–19 (746,005) | 20–29 (618,935) | 30–39 (412,302) | 40–49 (259,683) | 50–59 (147,204) | 60–69 (68,053) | 70–79 (29,359) | 80–89 (11,889) | 90–99 (4,425) | 100–124 (2,730) | 125–149 (380) | 150–199 (122) | | |
| CDR user activity (number of records per week) | 0 – 19 (687435) | $7.60 \cdot 10^{-6}$ (0.148) | $1.47 \cdot 10^{-4}$ (0.399) | $6.49 \cdot 10^{-4}$ (0.656) | $1.66 \cdot 10^{-3}$ (0.898) | $3.18 \cdot 10^{-3}$ (1.12) | $5.35 \cdot 10^{-3}$ (1.33) | $8.52 \cdot 10^{-3}$ (1.55) | 0.0128 (1.78) | 0.0182 (2.01) | 0.0279 (2.32) | 0.0446 (2.73) | 0.0911 (3.53) | 0.1527 (4.29) | $1.28 \cdot 10^{-3}$ | $1.84 \cdot 10^{-3}$ |
| | 20 – 29 (155136) | $2.44 \cdot 10^{-4}$ (0.473) | $4.83 \cdot 10^{-3}$ (1.28) | 0.0211 (2.11) | 0.0527 (2.9) | 0.0968 (3.61) | 0.1528 (4.29) | 0.2239 (5.02) | 0.3028 (5.76) | 0.3821 (6.48) | 0.4888 (7.5) | 0.6077 (8.81) | 0.7686 (11.4) | 0.8576 (13.9) | 0.0356 | 0.0515 |
| | 30 – 49 (244241) | $9.72 \cdot 10^{-4}$ (0.75) | 0.0193 (2.05) | 0.0813 (3.39) | 0.1862 (4.65) | 0.3071 (5.8) | 0.4276 (6.9) | 0.5442 (8.07) | 0.6424 (9.25) | 0.7186 (10.4) | 0.7971 (12) | 0.8637 (14.1) | 0.9312 (18.2) | 0.9565 (22.1) | 0.1063 | 0.1541 |
| | 50 – 69 (202655) | $3.08 \cdot 10^{-3}$ (1.1) | 0.0600 (3.03) | 0.2261 (5.04) | 0.4306 (6.93) | 0.5954 (8.66) | 0.7126 (10.3) | 0.7984 (12.1) | 0.8562 (13.8) | 0.8942 (15.5) | 0.9283 (17.9) | 0.9542 (21) | 0.9588 (27) | 0.9616 (33) | 0.2157 | 0.3123 |
| | 70 – 99 (256755) | $7.91 \cdot 10^{-3}$ (1.51) | 0.1432 (4.19) | 0.4364 (6.99) | 0.6676 (9.61) | 0.7967 (12) | 0.8684 (14.3) | 0.9133 (16.7) | 0.9406 (19.1) | 0.9562 (21.5) | 0.9578 (24.8) | 0.9598 (29.1) | 0.9637 (37.4) | 0.9675 (45.6) | 0.3296 | 0.4759 |
| | 100 – 149 (330530) | 0.0206 (2.09) | 0.3086 (5.81) | 0.6757 (9.72) | 0.8441 (13.4) | 0.9138 (16.7) | 0.9470 (19.9) | 0.9571 (23.3) | 0.9587 (26.7) | 0.9602 (30) | 0.9623 (34.6) | 0.9651 (40.5) | 0.9706 (52.3) | 0.9759 (63.7) | 0.4500 | 0.6454 |
| | 150 – 199 (237813) | 0.0446 (2.73) | 0.5001 (7.61) | 0.8246 (12.8) | 0.9244 (17.6) | 0.9565 (22) | 0.9584 (26.2) | 0.9605 (30.7) | 0.9626 (35.1) | 0.9646 (39.5) | 0.9674 (45.5) | 0.9711 (53.4) | 0.9783 (68.9) | 0.9853 (83.9) | 0.5415 | 0.7676 |
| | 200 – 249 (170137) | 0.0813 (3.39) | 0.6553 (9.43) | 0.8993 (15.8) | 0.9564 (21.8) | 0.9589 (27.3) | 0.9614 (32.5) | 0.9639 (38) | 0.9665 (43.6) | 0.9691 (49.1) | 0.9726 (56.6) | 0.9772 (66.4) | 0.9862 (85.8) | 0.9950 (105) | 0.6058 | 0.8445 |
| | 250 – 299 (123475) | 0.1189 (3.9) | 0.7439 (10.9) | 0.9319 (18.2) | 0.9579 (25.1) | 0.9609 (31.5) | 0.9637 (37.6) | 0.9667 (44) | 0.9697 (50.4) | 0.9726 (56.7) | 0.9767 (65.4) | 0.9820 (76.8) | 0.9924 (99.2) | 1.0000 (121) | 0.6438 | 0.8828 |
| | 300 – 399 (157376) | 0.1882 (4.67) | 0.8335 (13) | 0.9564 (21.9) | 0.9603 (30.2) | 0.9639 (37.9) | 0.9673 (45.2) | 0.9709 (52.9) | 0.9745 (60.7) | 0.9781 (68.3) | 0.9829 (78.8) | 0.9893 (92.5) | 1.0000 (119) | 1.0000 (146) | 0.6909 | 0.9196 |
| | 400 – 499 (89358) | 0.2720 (5.48) | 0.8897 (15.3) | 0.9582 (25.7) | 0.9627 (35.4) | 0.9669 (44.5) | 0.9710 (53.1) | 0.9752 (62.2) | 0.9795 (71.4) | 0.9837 (80.5) | 0.9895 (92.8) | 0.9970 (109) | 1.0000 (141) | 1.0000 (172) | 0.7308 | 0.9396 |
| | 500 – 749 (102698) | 0.3910 (6.56) | 0.9327 (18.3) | 0.9606 (30.8) | 0.9661 (42.6) | 0.9712 (53.6) | 0.9761 (64.2) | 0.9813 (75.2) | 0.9864 (86.3) | 0.9916 (97.3) | 0.9984 (112) | 1.0000 (131) | 1.0000 (169) | 1.0000 (207) | 0.7792 | 0.9559 |
| | 750 – 999 (39523) | 0.5121 (7.73) | 0.9562 (21.5) | 0.9631 (36.3) | 0.9696 (50.3) | 0.9758 (63.5) | 0.9817 (76.2) | 0.9879 (89.4) | 0.9940 (103) | 1.0000 (116) | 1.0000 (133) | 1.0000 (155) | 1.0000 (199) | 1.0000 (243) | 0.8241 | 0.9661 |
| | 1000 – 1999 (38714) | 0.6456 (9.3) | 0.9582 (25.8) | 0.9665 (43.5) | 0.9745 (60.6) | 0.9820 (76.7) | 0.9893 (92.5) | 0.9968 (109) | 1.0000 (125) | 1.0000 (141) | 1.0000 (161) | 1.0000 (187) | 1.0000 (238) | 1.0000 (291) | 0.8686 | 0.9701 |
| | 2000 – 9999 (8791) | 0.7960 (12) | 0.9618 (33.4) | 0.9727 (56.7) | 0.9832 (79.3) | 0.9933 (101) | 1.0000 (122) | 1.0000 (144) | 1.0000 (165) | 1.0000 (186) | 1.0000 (211) | 1.0000 (243) | 1.0000 (308) | 1.0000 (378) | 0.9200 | 0.9765 |
| | average | 0.0686 | 0.3053 | 0.4527 | 0.5323 | 0.5807 | 0.6144 | 0.6416 | 0.6632 | 0.6807 | 0.6998 | 0.7198 | 0.7492 | 0.7738 | 0.3336 | |
| | limited average (users with between 20 – 1999 CDR records / week) | 0.0876 | 0.4003 | 0.5952 | 0.7003 | 0.7638 | 0.8077 | 0.8427 | 0.8700 | 0.8914 | 0.9135 | 0.9347 | 0.9588 | 0.9715 | | 0.5968 |

Table S4: **Success ratios for different groups of users by activity**, extrapolated for four weeks. Numbers in the table correspond to the success ratios in that group. Numbers in parentheses below the success ratios are the expected number of matches in that group. Numbers in parentheses below the user activities are the number of people in that group. The averages are calculated by weighting by the number of users. The last row and column displays averages calculated with leaving out casual users with very low activity (for whom this matching procedure is not expected to be possible), and users with unrealistically high activiy.



\end{document}